\documentclass[a4paper,fleqn]{cas-dc}

\usepackage[numbers]{natbib}
\usepackage{subfig}
\usepackage{float}
\usepackage{array}   
\usepackage[linesnumbered,ruled,vlined]{algorithm2e}
\usepackage{amsmath}
\usepackage{xcolor}

\begin{document}
\def\floatpagepagefraction{1}
\def\textpagefraction{.01}
\shorttitle{Yunsheng~Wang et~al. Expert Systems with Applications}
\shortauthors{Wang et~al.}

\title [mode = title]{AMB-DSGDN: Adaptive Modality-Balanced Dynamic Semantic Graph Differential Network for Multimodal Emotion Recognition}

\tnotetext[1]{Our code is publicly available at \href{https://github.com/wys-ljq/AMB-DSGDN}{https://github.com/wys-ljq/AMB-DSGDN}.}

\tnotetext[1]{Email: 20234261@csuft.edu.cn, shouyuntao@stu.xjtu.edu.cn, tanyilong@csuft.edu.cn, aiwei@hnu.edu.cn, mengtao@hnu.edu.cn, lik@newpaltz.edu}

\author[1]{Yunsheng Wang}

\author[1]{Yuntao Shou}

\author[1]{Yilong Tan}

\author[1]{Wei Ai}

\author[1]{Tao Meng}
\cormark[1]

\author[2]{Keqin Li}

\cortext[1]{Corresponding author}

\address[1]{organization={College of Computer and Mathematics, Central South University of Forestry and Technology},
                postcode={410004},
                city={ Hunan, Changsha},
                country={China}}

\address[2]{organization={Department of Computer Science, State University of New York}, 
                city={New Paltz, New York},
                postcode={12561},
                country={USA}}

\begin{abstract}
 Multimodal dialogue emotion recognition captures emotional cues by fusing text, visual, and audio modalities. However, existing approaches still suffer from notable limitations in modeling emotional dependencies and learning multimodal representations. On the one hand, they are unable to effectively filter out redundant or noisy signals within multimodal features, which hinders the accurate capture of the dynamic evolution of emotional states across and within speakers. On the other hand, during multimodal feature learning, dominant modalities (e.g., textual cues) tend to overwhelm the fusion process, thereby suppressing the complementary contributions of non-dominant modalities such as speech and vision, ultimately constraining the overall recognition performance. To address these challenges, we propose an Adaptive Modality-Balanced Dynamic Semantic Graph Differential Network (AMB-DSGDN). Concretely, we first construct modality-specific subgraphs for text, speech, and vision, where each modality contains intra-speaker and inter-speaker graphs to capture both self-continuity and cross-speaker emotional dependencies. On top of these subgraphs, we introduce a differential graph attention mechanism, which computes the discrepancy between two sets of attention maps. By explicitly contrasting these attention distributions, the mechanism cancels out shared noise patterns while retaining modality-specific and context-relevant signals, thereby yielding purer and more discriminative emotional representations. In addition, we design an adaptive modality balancing mechanism, which estimates a dropout probability for each modality according to its relative contribution in emotion modeling. This mechanism randomly discards a portion of features from dominant modalities to suppress their overwhelming influence, while proportionally rescaling the preserved features based on the dropout probability to maintain overall information balance. Extensive experiments on IEMOCAP and MELD datasets validate that AMB-DSGDN significantly outperforms state-of-the-art baselines, demonstrating its effectiveness and robustness in multimodal conversational emotion recognition.
\end{abstract}

\begin{keywords}
\sep Multimodal Emotion Recognition  
\sep Adaptive Modality Balancing  
\sep Speaker Semantic Graph
\sep Graph Neural Networks
\end{keywords}

\maketitle

\section{Introduction}
Dialogue emotion recognition is a key task in human-computer interaction, natural language processing, and affective computing, aiming to accurately identify speakers' emotional states in multi-party dialogues to enhance the performance of intelligent systems in applications such as social robots, virtual assistants, mental health monitoring, and customer service  \cite{wu2025multimodal}, \cite{poria2018meld}, \cite{ai2025revisiting}, \cite{shou2024adversarial}. With the rapid development of artificial intelligence, dialogue emotion recognition has evolved from unimodal to multimodal approaches, integrating text, visual, and audio sources to capture the complexity and multidimensional features of human emotional expression \cite{wu2025multi}, \cite{tu2024multimodal}, \cite{shou2024low}, \cite{meng2024deep}. For instance, in everyday conversations, emotions are conveyed not only through verbal content but also through facial expressions and vocal tones, where the complementarity of these modalities helps models better understand emotional dynamics \cite{sun2025dialoguemllm}, \cite{shou2026comprehensive}, \cite{shou2022conversational}.

\begin{figure}[t]
	\centering
	\includegraphics[width=0.99\linewidth]{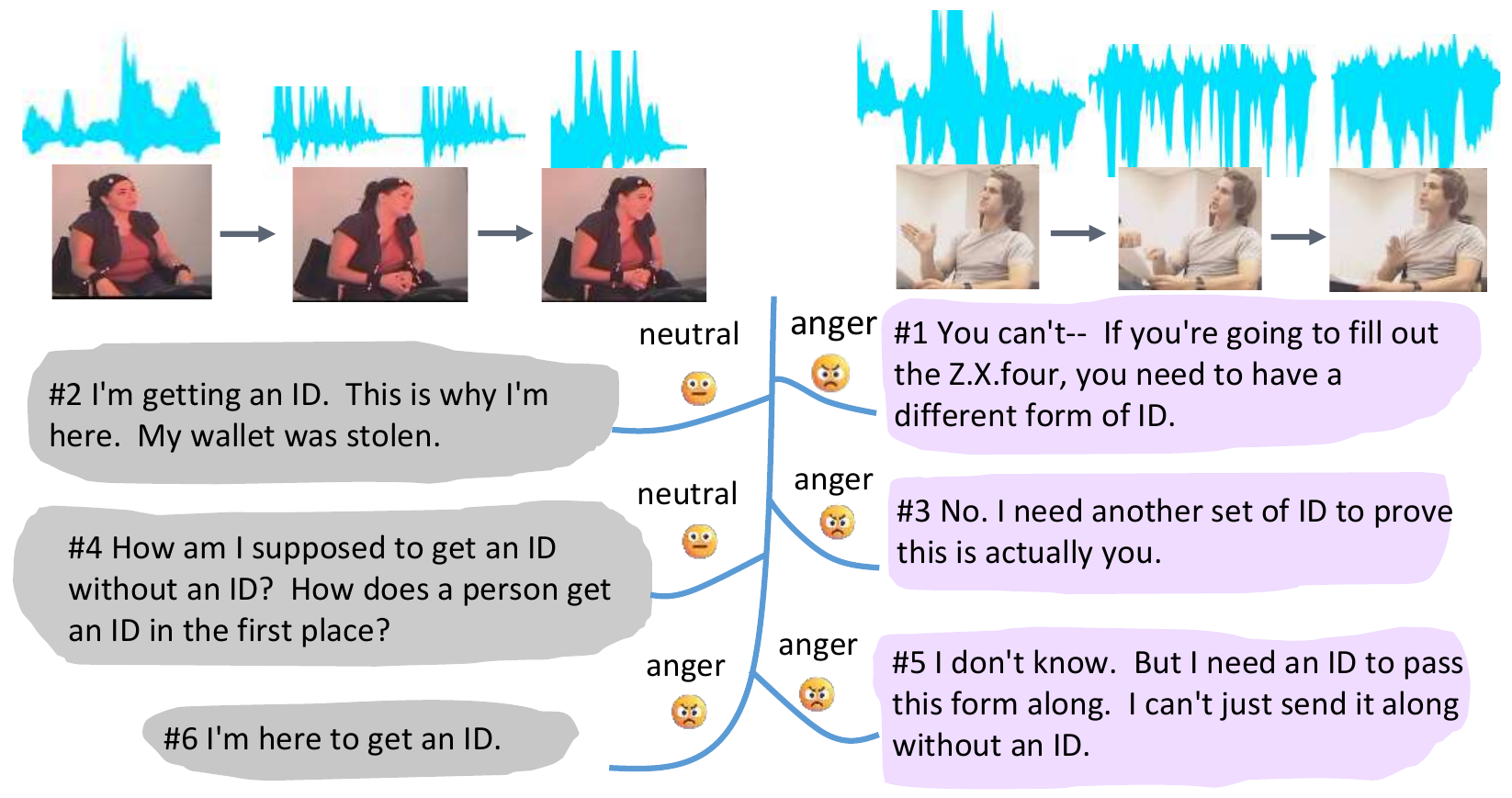}
	\caption{An authentic and representative segment illustrating the dynamic evolution of dialogue from the IEMOCAP dataset (Ses01F\_impro01).}
	\label{fig1}
\end{figure}
It is worth further attention that emotional states in dialogues evolve continuously with the interaction process, exhibiting significant dynamic characteristics \cite{fu2025feature}, \cite{shou2025masked}, \cite{meng2024multi}. In real scenarios, individuals may sustain prior emotional states or adjust instantly due to others' influences, reflecting intra-speaker continuity and inter-speaker interactivity \cite{mehrez2025multimodal}, \cite{shou2026graph}, \cite{shou2024efficient}. Meanwhile, the importance of different modalities fluctuates dynamically during the dialogue, affecting the model's ability to capture emotional changes \cite{lian2023survey}, \cite{shou2025spegcl}, \cite{ai2026paradigm}. Fig. \ref{fig1} illustrates a typical dialogue segment. The male, as a staff member, exhibits a stern attitude from the beginning, with his anger primarily manifested through tone and facial expressions, persisting in subsequent exchanges, reflecting the continuity of intra-speaker dependency. The female initially makes a request in a natural tone, with the text modality providing rational semantic information, but as the male's sternness continues, she shifts to anger, with changes in vocal intensity and facial expressions becoming more prominent, demonstrating sensitivity to the other's emotions. This process indicates that the female's emotional changes rely more on inter-speaker emotional influences, highlighting the dynamic contagion of cross-speaker dependencies. At the same time, this segment reveals the phased dominance of different modalities in the dialogue process: the text modality typically provides clearer semantic information in the early stages, while audio and visual modalities play a larger role when emotional changes are evident.

Unlike traditional static feature modeling, multimodal dialogue emotion recognition tasks require capturing the dynamic evolution of emotion dependencies over time \cite{zhao2025cross}, \cite{zhao2025enhanced}, \cite{cheng2024emotion}. If only static graph structures are employed for modeling, emotion dependencies are often reduced to fixed relational patterns, making it difficult to capture the dynamic changes in emotions driven by contextual variations, thus limiting the ability to model complex dialogue scenarios  \cite{farhadipour2025multimodal}, \cite{shou2025contrastive}, \cite{shou2025revisiting}, \cite{shou2025dynamic}. Meanwhile, multimodal features commonly contain redundant and shared noise, which, if indiscriminately incorporated into the modeling process, can obscure effective emotional signals and reduce the discriminative power of representations \cite{fan2024learning}, \cite{shou2025gsdnet}, \cite{shou2023graphunet}. Additionally, this task must address the fluctuating contributions of different modalities throughout the dialogue process. If the model overlooks the dynamic differences between modalities, it is likely to result in dominant modalities overly influencing the outcome, while weaker modalities fail to contribute effectively, ultimately degrading overall recognition performance \cite{shou2024dynamic}, \cite{shou2025graph}, \cite{shou2025multimodal}. Therefore, to effectively capture the continuously evolving emotional states in dialogues, it is essential to achieve dynamic modeling of emotion dependencies, effective suppression of noise in modeled features, and adaptive regulation of modality contributions.

To this end, researchers in multimodal dialogue emotion recognition (MERC) have proposed various modeling approaches, mainly divided into two directions: sequence structure methods rely on recurrent or memory networks to depict temporal dependencies, capable of capturing local continuity but struggling to cover long-distance inter-speaker interactions \cite{majumder2019dialoguernn}, \cite{shou2025graph}, \cite{shou2025cilf}; graph structure methods build dialogue graphs to model intra- and inter-speaker dependencies but mostly use static edge weights, ignoring the dynamic changes of dependency relationships with time and context \cite{ghosal2019dialoguegcn}. At the same time, for the fluctuating contributions of different modalities in the dialogue process, existing methods lack effective balancing regulation mechanisms, easily causing dominant modalities to prevail and weaker modalities to be diminished \cite{guo2024speaker}.

To address the aforementioned limitations, this paper proposes a Dynamic Semantic Graph Differential Network with Adaptive Modal Balancing. Specifically, for modeling emotion dependencies, we construct modality-specific subgraphs for text, speech, and visual modalities to capture both inter-speaker interactions and intra-speaker emotional evolution. On top of these subgraphs, we design a graph differential attention mechanism. This mechanism first projects utterance features into a unified representation space and computes attention distributions between nodes using left and right linear transformations, while explicitly modeling inter-speaker interactions by incorporating relational embeddings. Subsequently, a differential operation is performed on the two attention distributions, canceling out their overlapping components and retaining only the differential parts. This approach effectively removes shared noise present across different modalities while emphasizing modality-specific and contextually relevant dependency signals. Through this differential modeling strategy, the model not only filters out redundant information but also captures the dynamic dependencies evolving with context in dialogues, resulting in purer and more discriminative emotional representations. For modality regulation, we introduce an adaptive modal dropout mechanism. The model calculates dropout probabilities based on the relative performance of each modality in emotion recognition and randomly discards a portion of the dominant modality’s features through probabilistic sampling, while scaling the retained features to ensure the stability of the overall information content. This strategy effectively mitigates modality imbalance, preventing any single modality from overly dominating the fusion process. Through the synergistic interaction of these two mechanisms, the model can dynamically capture emotional evolution in dialogues. The main contributions of this paper are as follows:

\begin{itemize}
	\item \textcolor{black}{
We propose AMB-DSGDN, which explicitly constructs modality-specific subgraphs to model both intra-speaker and inter-speaker emotional dependencies. By jointly integrating differential graph attention and an adaptive modality balancing mechanism, the model effectively captures dynamic emotional variations in conversations while alleviating noise interference and modality imbalance, enhancing the discriminability of emotional representations.
    }
    
	\item 
   
We design a differential graph attention mechanism that computes discrepancies between paired attention maps on modality-specific subgraphs. Through differential contrast, this mechanism suppresses shared noise and highlights modality-specific and context-relevant information, improving dynamic emotion modeling capability.
    
	\item 
    
We further introduce an adaptive dropout-based modality balancing mechanism, which dynamically identifies the dominant modality and randomly drops part of its features, while proportionally rescaling the retained features, thereby alleviating the impact of single-modality dominance and enabling balanced multimodal information fusion.

    \item
    
    Extensive experiments on the IEMOCAP and MELD datasets demonstrate the superior performance and robustness of AMB-DSGDN.
\end{itemize}


\section{Related Works}

This section reviews recent advances in multimodal learning, dynamic emotional dependencies, and modal imbalance learning, highlighting key challenges and research directions for improving multimodal conversational emotion recognition.

\subsection{Multimodal Learning}

In recent years, multimodal learning has made significant progress in fields such as computer vision, natural language processing, and speech analysis. Researchers have proposed various methods to enhance multi-source information fusion and improve model generalization performance. In multimodal emotion recognition tasks, the NORM-TR model \cite{liu2024noise} effectively captures long-range dependencies between modalities and improves the model's robustness to noise and computational efficiency through a noise-robust feature extractor and noise-aware learning scheme combined with a Transformer fusion mechanism. The AffectGPT model \cite{lian2025affectgpt} uses a pre-fusion operation and multimodal large language model architecture, placing cross-modal interactions outside the LLM to capture fine-grained emotions from text, audio, and visual inputs. The HKD-MER model \cite{sun2024muti} enhances feature balance and discriminative ability by transferring dominant modality knowledge to other modalities through hierarchical knowledge distillation. The MERBench model \cite{lian2024merbench} achieves fine-grained modeling of modality heterogeneity and cross-modal interactions using an attention-based fusion framework. Overall, these methods have achieved positive results in fusion effects and generalization capabilities. However, existing research often fails to fully consider the dynamic differences in emotional expressions across modalities in dialogue scenarios, cannot effectively filter redundant noise in multimodal features, and the dominant modality (such as text) overly dominates the fusion process, suppressing the complementary role of non-dominant modalities, thereby limiting overall recognition performance. To this end, this paper constructs modality-specific subgraphs for dynamic emotion modeling, introduces differential graph attention to offset shared noise, and adopts an adaptive dropout strategy to adjust modality contributions, thereby improving joint learning in multimodal models.

\subsection{Dynamic Emotional Dependencies}

In the field of emotion recognition, recent research has proposed various methods to better capture dynamic emotional dependencies, mainly including two categories: recurrent neural network (RNN)-based and graph convolutional network (GCN)-based models. RNN-based methods (such as the optimized RNN model \cite{reddy2025optimized}) can handle sequential data and model temporal dependencies, but are prone to gradient vanishing or explosion in long sequences, limiting the expression of complex emotional dynamics. To enhance modeling capabilities, researchers use GCN to handle emotional association structures, suitable for dialogue scenarios with multi-turn interactions. DER-GCN \cite{ai2024gcn} strengthens inter-speaker dependency modeling by fusing dialogue-aware and event-aware information; SERC-GCN \cite{chandola2024serc} captures changes in speaker emotional states to better capture emotional evolution. Some studies also attempt to fuse RNN and GCN, such as GCN-LSTM \cite{kong2024emotional}, to balance temporal and structural information. Additionally, DEDNet \cite{wang2024dynamic} models inter- and intra-speaker emotional dependencies and uses interaction to capture emotion changes. Although these methods have made progress, most graph-based methods still rely on static edge weights, and noise in multimodal features weakens the characterization of dynamic emotional dependencies. Therefore, this paper constructs intra-speaker and inter-speaker subgraphs in each modality to express temporal and interactive relationships of emotions. At the same time, positive and negative branch differential attention is introduced in the graph, using the difference in their attention and amplifying stable and consistent emotional features to achieve more reliable dynamic emotion modeling.

\subsection{Modal Imbalance Learning}

Modality imbalance is one of the core challenges in multimodal learning in recent years. Different modalities differ in data quality, information density, and availability, leading to some modalities dominating the fusion while others' contributions are weakened. To address this issue, researchers have proposed various strategies. Wei et al. \cite{wei2024enhancing} designed a dynamic modality assignment framework that adaptively adjusts weights for each modality, enhancing the contribution of secondary modalities and alleviating single-modality dominance. Wang et al. \cite{wang2025mitigating} proposed an adversarial modality balancing method that enhances underrepresented modality features through quantity-quality reweighting, improving fusion effects. The MPLMM model proposed by Guo et al. \cite{guo2024multimodal} dynamically suppresses noisy modalities in emotion recognition while retaining key information. The MER framework proposed by Lian et al. \cite{lian2024mer} combines modality robustness with semi-supervised learning to enhance the discriminative ability of secondary modalities. Chen et al. \cite{chen2024efficiency} use multimodal knowledge distillation in a teacher-student architecture to achieve cross-modal knowledge transfer, alleviating performance degradation caused by modality imbalance. Nevertheless, existing methods do not fully consider the fluctuations in modality contributions with contextual changes in dynamic dialogues, leading to greater model fluctuations under extreme imbalances. To this end, this paper proposes an adaptive modality balancing strategy that dynamically adjusts feature sampling by evaluating each modality's performance in the current batch and proportionally amplifies retained features, thereby moderately suppressing features of dominant modalities when they are too strong while enhancing secondary modality contributions to achieve balanced fusion of multimodal features.

\section{Task Definition}

Multimodal Emotion Recognition in Conversation aims to identify the emotion category of each utterance in a dialogue sequence. The input to this task is a dialogue sequence comprising multiple utterances, where each utterance includes features from text, visual, and audio modalities, along with speaker information and contextual relationships. Formally, given a dialogue sequence \(\mathcal{D} = \{u_1, u_2, \dots, u_N\}\), where \(N\) is the number of utterances, each utterance \(u_i\) consists of its text feature \(\mathbf{t}_i \in \mathbb{R}^{d_t}\), visual feature \(\mathbf{v}_i \in \mathbb{R}^{d_v}\), audio feature \(\mathbf{a}_i \in \mathbb{R}^{d_a}\), speaker identifier \(s_i\), and emotion label \(y_i \in \{1, 2, \dots, C\}\) (\(C\) is the number of emotion categories). The model's objective is to learn a mapping function \(f: \mathcal{D} \rightarrow \{y_1, y_2, \dots, y_N\}\) to maximize prediction accuracy and weighted F1 score. This task incorporates multimodal fusion, speaker dependencies, and contextual relationships, making it suitable for datasets such as IEMOCAP and MELD.

\section{Methodology}
In this section, we introduce the proposed multimodal emotion recognition method for conversations. The method encompasses six key components:  model architecture, utterance-level encoder, relational subgraph construction, differential attention graph convolutional network, dynamic modality balancing mechanism, and emotion classifier.

\begin{figure*}[t]
	\centering
	\includegraphics[width=0.98\linewidth]{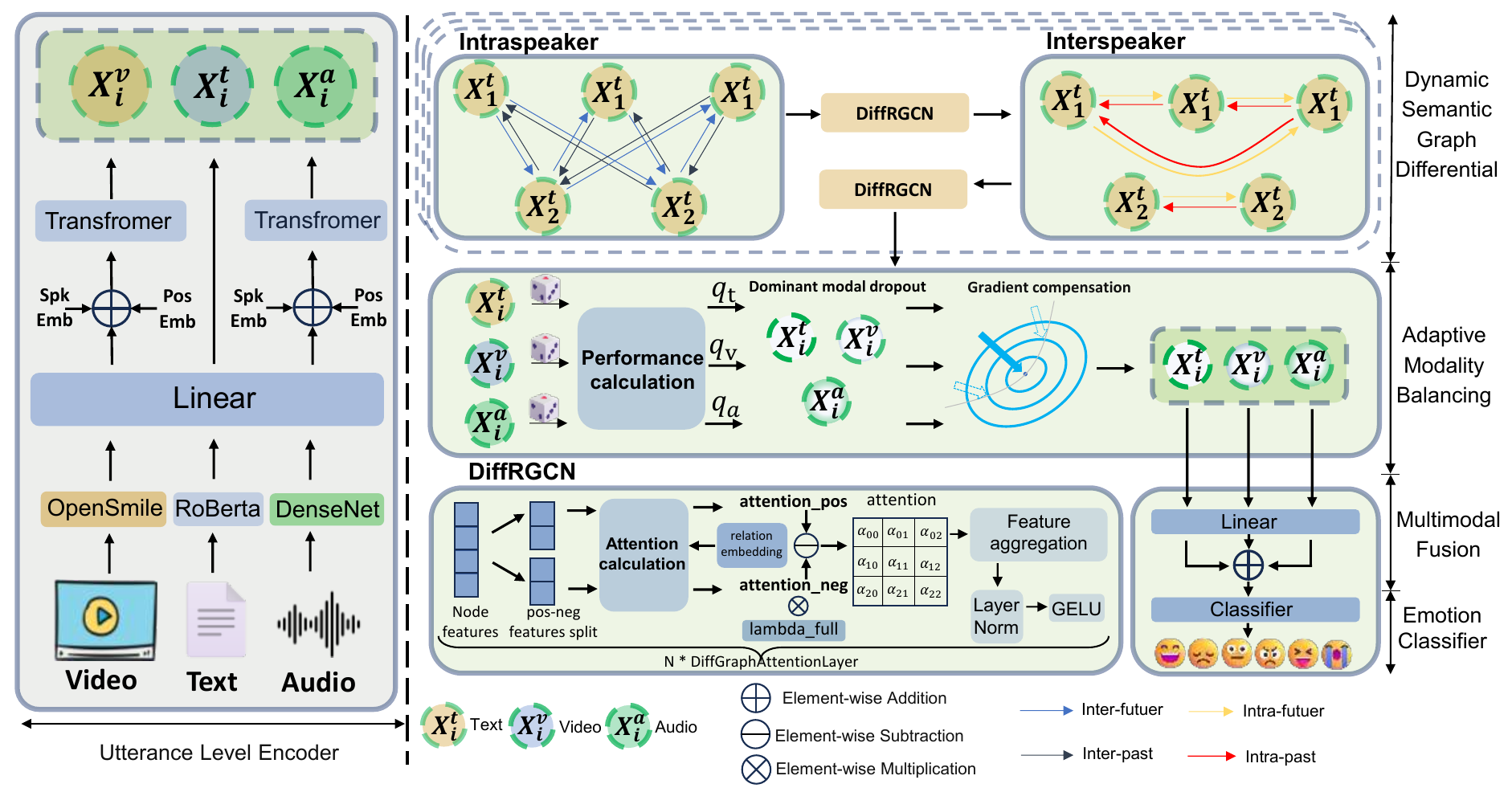}
	\caption{This architecture includes four core modules: first, the utterance-level encoder extracts unimodal features through OpenSmile (audio), RoBERTa (text), DenseNet (video), and after Transformer combining speaker embedding (Spk Emb) and position embedding (Pos Emb) encoding, obtains the text feature \(\mathbf{x}_i^t\), video feature \(\mathbf{x}_i^t\), audio feature \(\mathbf{x}_i^t\) for the \(i\)-th utterance; second, the differential graph attention module constructs subgraphs including "intra-speaker subgraph (Intraspeaker)" and "inter-speaker subgraph (Interspeaker)" for each modality, computes differences between two groups of attention distributions through differential graph attention, eliminates cross-modal shared noise and retains modality-specific emotional signals; then, the adaptive modality balancing module computes dropout probabilities (\(q_t/q_v/q_a\)) for each modality based on batch-level performance, performs dynamic dropout on dominant modalities, while scaling retained features through gradient compensation to maintain information balance; finally, through multimodal fusion and classification module, fuses the balanced features via linear layers and inputs into the classifier to obtain emotion recognition results.}
	\label{fig2}
\end{figure*}

\subsection{Model Architecture}

AMB-DSGDN combines differential attention graph convolutional networks with dynamic modality balancing mechanisms. Fig. \ref{fig2} shows the overall framework of the model, mainly consisting of five core components: utterance-level encoder, relational subgraph construction, differential attention graph convolutional network, dynamic modality balancing mechanism, and emotion classifier. First, text, audio, and visual features are mapped to unified dimensional hidden representations through respective linear transformations. Subsequently, a Transformer encoder incorporating speaker embeddings is introduced for contextual modeling to capture temporal dependencies between utterances. On this basis, the model constructs intra-speaker and inter-speaker relational subgraphs and uses differential attention graph convolution mechanisms to model emotional dependencies. This mechanism suppresses shared redundancies and noise patterns by modeling differences in node attention distributions in the same modality subgraph, thereby highlighting true emotional dependency signals. This design helps more accurately characterize dynamic emotional changes in dialogues. In terms of modality regulation, the adaptive modality dropout mechanism generates dropout probabilities based on each modality's contribution to emotion modeling and scales compensated retained features to achieve dynamic weight adjustment across modalities. Finally, the fused multimodal features are fed into the emotion classifier for prediction, while auxiliary losses are introduced to enhance the robustness of unimodal emotional representations.

\subsection{Utterance-Level Encoder}

The utterance-level encoder is used to perform feature extraction and fusion on multimodal inputs (text, visual, and audio) to generate semantic representations for each utterance in the dialogue. This module consists of modality feature projection, position encoding and speaker embedding, and Transformer encoding layers, ultimately outputting unified representations for the three modalities.

For each utterance in the dialogue, initial features are first extracted from pre-trained models: text modality uses RoBERTa \cite{liu2019roberta}, visual modality uses DenseNet \cite{huang2017densely}, audio modality uses OpenSmile \cite{eyben2010opensmile}. Subsequently, linear mappings project different modality features to unified hidden dimensions:
\begin{equation}
\mathbf{x}_i^t = W_t \cdot \mathbf{f}_i^t +\mathbf{b}_t,\quad \mathbf{x}_i^v = W_v \cdot \mathbf{f}_i^v + \mathbf{b}_v,\quad \mathbf{x}_i^a = W_a \cdot \mathbf{f}_i^a + \mathbf{b}_a,
\end{equation}
where \(\mathbf{f}_i^t\),\(\mathbf{f}_i^v\),\(\mathbf{f}_i^a\) represent the original text, visual, and audio features of the i-th utterance, respectively, \(W_t \in \mathbb{R}^{d_h \times D_t}\),\(W_v \in \mathbb{R}^{d_h \times D_v}\),\(W_a \in \mathbb{R}^{d_h \times D_a}\) are projection matrices, \(\mathbf{b}_t\),\(\mathbf{b}_v\),\(\mathbf{b}_a\) are bias terms.

To introduce sequence position information and speaker identity information, the model adds position encoding and speaker embedding. Position encoding uses sine-cosine form:
\begin{equation}
\begin{split}
PE(pos, 2j) &= \sin\left(\frac{pos}{10000^{2j/d_h}}\right), \\ 
PE(pos, 2j+1) &= \cos\left(\frac{pos}{10000^{2j/d_h}}\right)  
\end{split}
\end{equation}
where \(pos\) represents the position of the utterance in the sequence, \(j\) is the dimension index. Speaker embedding is generated through an embedding layer:
\(\mathbf{s}_i = E_s(\text{spk}_i)\),where \(E_s \in \mathbb{R}^{(n_s + 1) \times d_h}\),\(n_s\) is the number of speakers.

For audio and visual modalities, add the projected features with position encoding and speaker embedding:
\begin{equation}
\mathbf{x}_i^a = \mathbf{x}_i^a + PE(i) + \mathbf{s}_i, \quad \mathbf{x}_i^v = \mathbf{x}_i^v + PE(i) + \mathbf{s}_i.
\end{equation}

Subsequently, audio and visual modalities are input into independent Transformer encoders to model contextual dependencies. Each encoder consists of a single-layer multi-head self-attention and position-wise feed-forward network. The self-attention mechanism is defined as:
\begin{equation}
\text{Attention}(Q, K, V) = \text{softmax}\left(\frac{QK^T}{\sqrt{d_k}}\right)V,
\end{equation}
where \(Q = W_q \mathbf{x}\),\(K = W_k \mathbf{x}\),\(V = W_v \mathbf{x}\),\(d_k = d_h / h\),\(h\) is the number of attention heads. The position-wise feed-forward network is defined as:
\begin{equation}
FFN(\mathbf{x}) = W_2 \cdot \text{GELU}(W_1 \cdot LN(\mathbf{x})) + \mathbf{x},
\end{equation}
where \(LN\) represents layer normalization.

Finally, the contextual representations of audio and visual modalities are uniformly represented as:
\begin{equation}
\mathbf{x}_i^m, \mathbf{A}_{attn}^m = \text{TransformerEncoder}(\mathbf{x}^m, \mathbf{mask}, \mathbf{s}), \quad m \in \{a, v\},
\end{equation}
where \(\mathbf{A}_{attn}^m\) is the attention matrix reflecting dependencies between utterances, \(\mathbf{mask}\) is the utterance mask, \(\mathbf{s}\) represents sequence information.

\subsection{Relational Subgraph Construction}

In dialogues, each utterance \(u_i\) is represented as graph node \(v_i\), with multimodal features \(\mathbf{x}_i^m, m \in \{t,v,a\}\), corresponding to text, visual, and audio modalities respectively.

\subsection{Graph Construction and Representation}

A complete dialogue is represented as \(U = \{u_1, u_2, \ldots, u_{N_b}\}\), and modeled as a directed graph \(\mathcal{G} = (\mathcal{V}, \mathcal{E}, \mathcal{R}, \mathcal{W})\), where each utterance \(u_i\) corresponds to node \(v_i \in \mathcal{V}\). Node features consist of multimodal representations \((\mathbf{x}_i^t, \mathbf{x}_i^v, \mathbf{x}_i^a)\), corresponding to text, visual, and audio modalities. To model emotional dependencies, the model constructs two relational subgraphs: intra-speaker subgraph \(\mathbf{Adj}_s\) and inter-speaker subgraph \(\mathbf{Adj}_c\), characterizing temporal evolution within the same speaker and interactive relationships between different speakers respectively. Edge set \(\mathcal{E}\) represents dependencies between utterances, edge types \(\mathcal{R}\) encode temporal and emotional relationships into five categories: self-past, self-future, inter-speaker past, inter-speaker future, and self-loop. Edge weights \(\mathcal{W}\) represent interaction strength or proximity. Adjacency range is constrained by window size \(w=5\) to reduce long-distance noise and computational complexity. During batch processing, dialogue maximum length is set to \(L=\max_b N_b\), with padded positions set to zero. The two subgraphs are stacked as \(\mathbf{Adj}_s\) and \(\mathbf{Adj}_c\), providing structural information for subsequent graph attention modeling.

The intra-speaker subgraph \(\mathbf{Adj}_s\) only connects utterances from the same speaker, used to model internal consistency and temporal evolution:
\begin{equation}
(\mathbf{Adj}_s)_{i,j} =
\begin{cases}
1 & \text{if } i = j \land s_i = s_j, \\
2 & \text{if } i > j \land |i-j| \leq w \land s_i = s_j \quad (\text{intra-past}), \\
3 & \text{if } i < j \land |i-j| \leq w \land s_i = s_j \quad (\text{intra-future}), \\
0 & \text{otherwise}.
\end{cases}
\end{equation}

This design highlights the cumulative impact of past utterances on current emotions while introducing forward-looking context from future utterances.

The inter-speaker subgraph \(\mathbf{Adj}_c\) is used to model dynamic interactive relationships between different speakers:
\begin{equation}
(\mathbf{Adj}_c)_{i,j} =
\begin{cases}
1 & \text{if } i = j, \\
4 & \text{if } i < j \land |i-j| \leq w \land s_i \neq s_j \quad (\text{inter-future}), \\
5 & \text{if } i > j \land |i-j| \leq w \land s_i \neq s_j \quad (\text{inter-past}), \\
0 & \text{otherwise}.
\end{cases}
\end{equation}

This subgraph characterizes responses, conflicts, and collaborations in inter-speaker emotional dynamics.

The two relational subgraphs are finally represented as \(\mathbf{Adj}_s, \mathbf{Adj}_c \in \mathbb{Z}^{B \times L \times L}\),where \(r_{i,j} \in \mathcal{R}\),and \(\mathbf{Adj}_{i,j} = \text{id}(r_{i,j})\). The separated subgraph structures support hierarchical modeling, first aggregating global interaction information across speakers, then refining single-speaker internal representations, thereby enhancing modeling capabilities in multi-speaker scenarios.

The above relational subgraphs are shared across different modalities but act on their respective modality representations. Specifically, text, visual, and audio modalities use independent differential graph attention networks and model on the same \(\mathbf{Adj}_s\) and \(\mathbf{Adj}_c\). Taking text modality as an example, inter-speaker GAT (`gatTer`) first acts on \(\mathbf{Adj}_c\), followed by intra-speaker GAT (`gatT`) on \(\mathbf{Adj}_s\). Audio and visual modalities use `gatAer` / `gatA` and `gatVer` / `gatV` respectively. This design allows each modality to independently learn relation representations while sharing dialogue structures.

\subsection{Differential Attention Graph Convolutional Network}

To fuse structural information in relation subgraphs, we propose the Differential Relational Graph Convolutional Network (DiffRGCN). This network introduces a differential attention mechanism based on GAT to model relation types and feature differences. DiffRGCN consists of multi-head attention layers and an output layer, where each attention head implicitly characterizes enhancement and suppression relationships between nodes through the differential mechanism and enhances semantic awareness by combining relation labels. This network can be applied independently to different modalities and sequentially acts on inter-speaker subgraphs and intra-speaker subgraphs, first aggregating inter-speaker interactions, then refining intra-speaker representations, with the specific process shown in Alg. \ref{alg1}.

\begin{algorithm}[!t] 
\caption{Differential Graph Attention Layer}
\label{alg1}
\KwIn{Node features $\mathbf{H}^m$, adjacency matrix $\mathbf{Adj}$, layer depth $d$, relation-aware flag}

Project node features using Eqn.~(9) to obtain $\mathbf{W}\mathbf{h}$\;
Split $\mathbf{W}\mathbf{h}$ into positive and negative branches\;

Generate $\mathbf{q}^{\text{pos}}, \mathbf{k}^{\text{pos}}, \mathbf{v}^{\text{pos}}$ and 
$\mathbf{q}^{\text{neg}}, \mathbf{k}^{\text{neg}}, \mathbf{v}^{\text{neg}}$ using Eqn.~(10) and Eqn.~(11)\;

\If{relation-aware enabled}{
    Embed adjacency relations into attention scores\;
}

Compute raw attention scores $\mathbf{e}^{\text{pos}}$ and $\mathbf{e}^{\text{neg}}$  using Eqn.~(12)\;
Apply activation, masking, and Softmax normalization to obtain 
$\boldsymbol{\alpha}^{\text{pos}}$ and $\boldsymbol{\alpha}^{\text{neg}}$\;

Compute differential coefficients $\lambda_1$, $\lambda_2$, and $\lambda_{\text{full}}$ 
using Eqn.~(13) and Eqn.~(14)\;

Fuse positive and negative attention weights to obtain final 
$\boldsymbol{\alpha}$ using Eqn.~(15)\;

Concatenate value vectors $\mathbf{v}^{\text{pos}}$ and $\mathbf{v}^{\text{neg}}$ 
to obtain $\mathbf{W}\mathbf{h}_v$ (Eqn.~(16))\;

Aggregate $\mathbf{W}\mathbf{h}_v$ using attention weights $\boldsymbol{\alpha}$ 
to obtain $\mathbf{h}'$ (Eqn.~(17))\;

Concatenate multi-head outputs to obtain final node representations 
$\mathbf{x}$ (Eqn.~(18))\;

\Return $\mathbf{x}$ and attention weights $\boldsymbol{\alpha}$\;
\end{algorithm}

The core of DiffRGCN is the differential graph attention layer. This layer divides the input features into positive and negative branches, modeling emotional enhancement and suppression signals respectively, and achieves differential aggregation through learnable lambda parameters. Specifically, the input features first undergo linear projection:
\begin{equation}
\mathbf{W}\mathbf{h} = \mathbf{H}^m \mathbf{W},
\end{equation}
where $\mathbf{H}^m \in \mathbb{R}^{B \times L \times d_{\mathrm{in}}}$, $\mathbf{W} \in \mathbb{R}^{d_{\mathrm{in}} \times d_{\mathrm{out}}}$, $B$ is the batch size, and $L$ is the number of nodes.

In the positive branch, the projected features are used to generate query and key vectors:
\begin{equation}
\mathbf{q}^{\text{pos}} = \mathbf{a}_{\text{left}}^{\text{pos}} \big( \mathbf{W}\mathbf{h}^{\text{pos}} \big),
\quad
\mathbf{k}^{\text{pos}} = \mathbf{a}_{\text{right}}^{\text{pos}} \big( \mathbf{W}\mathbf{h}^{\text{pos}} \big).
\end{equation}

The negative branch correspondingly generates $\mathbf{q}^{\text{neg}}$ and $\mathbf{k}^{\text{neg}}$.

The value vectors for positive and negative branches are defined as:
\begin{equation}
\mathbf{v}^{\text{pos}} = \mathbf{W}\mathbf{h}^{\text{pos}},
\quad
\mathbf{v}^{\text{neg}} = \mathbf{W}\mathbf{h}^{\text{neg}}.
\end{equation}

When relation awareness is enabled, the adjacency matrix is embedded as a relation representation $\mathbf{R} \in \mathbb{R}^{B \times L \times L \times d_r}$ and projected to obtain $\mathbf{rel}^{\text{pos}}$ and $\mathbf{rel}^{\text{neg}}$, which are added to the base attention scores:
\begin{equation}
\begin{split}
\mathbf{e}^{\text{pos}}_{i,j} &= \mathbf{q}^{\text{pos}}_i + \big( \mathbf{k}^{\text{pos}}_j \big)^\top + \mathbf{rel}^{\text{pos}}_{i,j}, \\
\mathbf{e}^{\text{neg}}_{i,j} &= \mathbf{q}^{\text{neg}}_i + \big( \mathbf{k}^{\text{neg}}_j \big)^\top + \mathbf{rel}^{\text{neg}}_{i,j}.
\end{split}
\end{equation}

The positive attention scores undergo LeakyReLU activation, mask invalid edges with $\mathbf{Adj} \le 0$, and are then normalized via Softmax to obtain positive attention weights $\boldsymbol{\alpha}^{\text{pos}}$. The negative branch follows the same process to obtain $\boldsymbol{\alpha}^{\text{neg}}$.

To balance positive and negative attention, differential lambda parameters are introduced. First, we compute:
\begin{align}
\begin{split}
\lambda_1 &= \exp \left( \sum_{d=1}^{D} \lambda_{\text{left},1}^{(d)} \cdot \lambda_{\text{right},1}^{(d)} \right), \\
\lambda_2 &= \exp \left( \sum_{d=1}^{D} \lambda_{\text{left},2}^{(d)} \cdot \lambda_{\text{right},2}^{(d)} \right),
\end{split}
\end{align}
where all parameters are initialized from a normal distribution with mean 0 and variance 0.1. The final differential coefficient is defined as:
\begin{equation}
\lambda_{\text{full}} = \lambda_1 - \lambda_2 + \lambda_{\text{init}},
\end{equation}
where $\lambda_{\text{init}} = 0.8 - 0.6 \exp(-0.3 \cdot \text{depth})$, and $\text{depth}$ denotes the layer depth.

The final attention weights are obtained through differential fusion:
\begin{equation}
\boldsymbol{\alpha} = \boldsymbol{\alpha}^{\text{pos}} - \lambda_{\text{full}} \cdot \boldsymbol{\alpha}^{\text{neg}}.
\end{equation}

This operation enhances effective emotional dependencies while suppressing noise interference.

Subsequently, the value vectors of the positive and negative branches are concatenated:
\begin{equation}
\mathbf{W}\mathbf{h}_v = \text{cat}(\mathbf{v}^{\text{pos}}, \mathbf{v}^{\text{neg}}) \in \mathbb{R}^{B \times L \times d_{\mathrm{out}}},
\end{equation}
and neighbor feature aggregation is completed under the attention weights:
\begin{equation}
\mathbf{h}' = \boldsymbol{\alpha} \mathbf{W}\mathbf{h}_v.
\end{equation}

DiffRGCN adopts multi-head differential attention to model multi-view relationships in parallel, with outputs concatenated as:
\begin{equation}
\mathbf{x} = \text{cat}\big([\text{att}_i(\mathbf{x}, \mathbf{Adj})]\big).
\end{equation}

The concatenated result undergoes dropout, an output attention layer, and a fully connected layer with residual connection, and final node representations are obtained via layer normalization. 

This design combines multi-head mechanisms with differential modeling, effectively enhancing the expressive power and stability of relation modeling while keeping the computational complexity $O(BLwd \cdot h + BL^2 / h)$ controllable.

\subsection{Adaptive Modality Balancing}

In multimodal learning, the contribution of each modality to emotion modeling is typically imbalanced, which may cause weak modalities to be overlooked during fusion. To address this issue, we propose a Dynamic Modality Balancing mechanism, which dynamically adjusts modality weights via an Adaptive Modality Dropout strategy. The detailed procedure is illustrated in Alg. \ref{Alg2}. To ensure timely responsiveness to changes in modality contributions, the modality performance metric \(p_m\), ratio parameter \(r_{m,j}\), and dropout probability \(q_m\) are updated in each training batch. This mechanism is only activated after the model enters a stable training phase. The determination of the stable training phase is based on the sensitivity experiment results of the warm-up epochs. As shown in the experiments in Section \ref{sec:sensitivity_warmup} the model performance stabilizes and reaches the optimal level after approximately 60 training epochs on the IEMOCAP dataset. Therefore, this paper sets the warm-up epoch to about 60 to avoid the impact of early parameter fluctuations on model convergence.

\begin{algorithm}[t]
\caption{Adaptive Modality Dropout}
\label{Alg2}
\KwIn{Modality feature list $\mathbf{F} = [\mathbf{T}, \mathbf{V}, \mathbf{A}]$, batch size $B$, base dropout probability $q_{\text{base}}$, scaling factor $\lambda$, labels $\mathbf{y}$, training mask $\mathbf{U}$}
\For{each training batch}{
    \tcp{Compute modality performance $p_m$}
    \For{each modality $m$}{
        $p_m = \text{F1}(\mathbf{F}_m, \mathbf{y}, \mathbf{U})$\;
    }

    \tcp{Compute relative performance ratios and dropout probabilities}
    \For{each modality $m$}{
        $r_{m,j} = \frac{p_m}{p_j + \epsilon} - 1, \forall j \neq m$\;
        $q_m = \text{clip}(q_{\text{base}} \cdot (1 + \lambda \cdot \text{Softmax}(\text{ReLU}(r_{m,j}))), 0, 1)$\;
    }

    \tcp{Apply dropout and scaling}
    Randomly decide whether to apply dropout\;
    \If{dropout is applied}{
        Sample Bernoulli mask $\mathbf{M} \sim 1 - q$\;
        Apply mask: $\mathbf{F}' = \mathbf{M} \odot \mathbf{F}$\;
        Scale by expectation: $\mathbf{F}'' = \mathbf{F}' / (1-\theta)$\;
        Keep instances with at least one modality: $\mathbf{F}''' = \mathbf{F}''[\mathbf{u}]$\;
    }
    \Else{
        $\mathbf{F}''' = \mathbf{F}, \mathbf{u} = \mathbf{1}$\;
    }
}
\Return Dropped and scaled modality features $\mathbf{F}'''$ and valid instance mask $\mathbf{u}$.
\end{algorithm}

First, we calculate the modality-level performance metric \(p_m\) for each instance. We use the weighted F1-score for measurement:
\(p_m = \text{F1}(\hat{y}_m, y, \mathbf{U})\), where \(\hat{y}_m = \arg\max(\mathbf{\hat{P}})\), \(y\) denotes the ground-truth label, and \(\mathbf{U}\) is the utterance mask used to filter padding positions. It is defined as:
\begin{equation}
p_m = \frac{\sum_{c=1}^C w_c \cdot \text{Precision}_c \cdot \text{Recall}_c}{\sum_{c=1}^C w_c},
\end{equation}
where \(w_c = \frac{N}{\sum_i \mathbb{I}(y_i = c)}\) is the inverse frequency weight of class \(c\). Precision and Recall are computed based on valid utterances. Specifically, we first flatten the logits and labels:
\(\mathbf{P}_m^{\flat} = \mathbf{P}_m[\mathbf{U} == 1]\), \(y^{\flat} = y[\mathbf{U} == 1]\), then calculate
\(\text{Precision}_c = \frac{\text{TP}_c}{\text{TP}_c + \text{FP}_c}\),
\(\text{Recall}_c = \frac{\text{TP}_c}{\text{TP}_c + \text{FN}_c}\).

Subsequently, we compute the relative performance differences between modalities. For modality \(m\), its performance ratio with other modalities is defined as:
\begin{equation}
r_{m,j} = \frac{p_m}{p_j + \epsilon} - 1, \quad \forall j \neq m,
\end{equation}
where \(\epsilon = 10^{-5}\) is a smoothing factor to avoid division by zero. Furthermore, for any modality \(m\), the dimension of its relative performance ratio vector \(\mathbf{r}_m\) is \(|\mathcal{M}| - 1\), where \(\mathcal{M} = \{t, v, a\}\). Each dimension of \(\mathbf{r}_m\) corresponds to the performance ratio \(r_{m,j}\) between modality \(m\) and another modality \(j (j \neq m)\), i.e.:
\begin{equation}
\mathbf{r}_m = [r_{m,j}]_{j \in \mathcal{M}, j \neq m}.
\end{equation}

In subsequent calculations, we do not explicitly filter positive values of \(\mathbf{r}_m\); instead, we directly apply the ReLU activation to the complete ratio vector to suppress negative terms, and perform Softmax normalization on the non-negative results to smooth the relative performance differences between modalities. Specifically, we construct the ratio vector \(\mathbf{r}_m\) by arranging all relative performance ratios \(r_{m,j} (j \in \mathcal{M}, j \neq m)\) in index order, first applying non-negative constraint:
\(\mathbf{r}_m^+ = \max(\mathbf{r}_m, 0)\), then performing Softmax normalization:
\begin{equation}
\hat{\mathbf{r}}_m = \frac{\exp(\mathbf{r}_m^+)}{\sum \exp(\mathbf{r}_m^+)},
\end{equation}
and finally obtaining the weighted average ratio:
\begin{equation}
\bar{r}_m = \frac{\sum \hat{\mathbf{r}}_m \odot \mathbf{r}_m}{|\mathbf{r}_m|}.
\end{equation}

Based on this ratio, the modality dropout probability is defined as:
\begin{equation}
q_m = q_{\text{base}} \cdot (1 + \lambda \cdot \bar{r}_m), \quad q_m = \text{clip}(q_m, 0, 1),
\end{equation}
where \(q_{\text{base}} = 0.3\) and \(\lambda = 0.9\) are hyperparameters. This design makes high-performance modalities have a higher dropout probability, thereby encouraging the model to focus on weaker modalities while avoiding extreme differences from dominating the training process.

When performing modality dropout, for the modality feature set
\(\mathbf{F} = [\mathbf{T}, \mathbf{V}, \mathbf{A}] \in \mathbb{R}^{M \times B \times N \times d_h}\) (where \(M = 3\) and \(B\) is the number of instances), we first decide whether to perform dropout with probability \(p_{\text{exe}} = 0.5\). If dropout is performed, we generate a Bernoulli mask:
\begin{equation}
\mathbf{M}_{m,b} \sim \text{Bernoulli}(1 - q_m),
\end{equation}
and apply it to the features:
\(\mathbf{F}' = \mathbf{M} \odot \mathbf{F}\).

Subsequently, we perform expectation compensation scaling via a custom Autograd function:
\begin{equation}
\mathbf{F}'' = \frac{\mathbf{F}'}{1 - \theta}, \quad \theta = \frac{\sum_m d_m q_m}{\sum_m d_m},
\end{equation}
where \(d_m = d_h\) is the modality dimension. Its forward propagation is linear scaling, and the backward propagation only passes gradients to the input:
\begin{equation}
\frac{\partial \mathbf{F}''}{\partial \mathbf{F}'} = \frac{1}{1 - \theta}, \quad
\frac{\partial \mathbf{F}''}{\partial \theta} = 0.
\end{equation}

Finally, we filter valid instances using the update flag \(\mathbf{u}_b = \sum_m \mathbf{M}_{m,b} > 0\), retaining only samples with at least one modality:
\(\mathbf{F}''' = \mathbf{F}''[\mathbf{u}]\), thus avoiding the empty representation problem caused by full modality dropout.

\subsection{Emotion Classifier}

After obtaining multimodal dynamic representations, we design a multimodal fusion and classification module to integrate text, visual, and audio features and generate emotion predictions. For the final representation of each modality, the model uses independent classification heads for projection. Each classification head consists of ReLU activation, Dropout layer, and linear mapping to map hidden representations to class space. Taking the text modality as an example, its classification process is defined as:
\begin{equation}
\mathbf{t} = \sigma(\text{Dropout}(\text{ReLU}(\mathbf{x}_i^t \mathbf{W}_t + \mathbf{b}_t))).
\end{equation}
where \(\mathbf{W}_t \in \mathbb{R}^{d_h \times C}\), \(\mathbf{b}_t \in \mathbb{R}^{C}\), \(\sigma\) represents the output after linear mapping. The classification heads \(\mathbf{v}\) and \(\mathbf{a}\) for visual and audio modalities are constructed in the same way.

Subsequently, fusion is completed by element-wise addition of each modality's logits:
\begin{equation}
\mathbf{O} = \mathbf{t} + \mathbf{v} + \mathbf{a},
\end{equation}
and log-softmax is used to compute the fused probability distribution:
\begin{equation}
\mathbf{P} = \log(\text{softmax}(\mathbf{O})), \quad \mathbf{\hat{P}} = \text{softmax}(\mathbf{O}).
\end{equation}

The predicted label is obtained by argmax of \(\mathbf{\hat{P}}\). To enhance unimodal representation capabilities, the model simultaneously computes unimodal log-softmax probabilities \(\mathbf{P}_t\), \(\mathbf{P}_v\) and \(\mathbf{P}_a\) for auxiliary optimization.

The model training uses cross-entropy loss and incorporates sequence masks to adapt to variable-length inputs. The fusion loss is defined as:
\begin{equation}
\mathcal{L}_{\text{fusion}} = \text{CE}(\mathbf{P}, \mathbf{y}, \mathbf{m}),
\end{equation}
where the cross-entropy form is:
\begin{equation}
\text{CE}(\mathbf{P}, \mathbf{y}, \mathbf{m}) = - \sum_{i=1}^{L} m_i \sum_{c=1}^{C} y_{i,c} \log P_{i,c}.
\end{equation}

Here, \(L\) represents sequence length, \(C\) represents number of classes, \(P_{i,c}\) is the predicted probability, \(y_{i,c}\) is the one-hot label, \(m_i \in {0,1}\) is the mask.
The final training objective consists of fusion loss and unimodal auxiliary losses:
\begin{equation}
\mathcal{L} = \mathcal{L}_{\text{fusion}} + \alpha_t \mathcal{L}_t + \alpha_a \mathcal{L}_a + \alpha_v \mathcal{L}_v,
\end{equation}
where \(\alpha_t = \mathcal{L}_t / 10\) etc. are adaptive weights to balance training contributions from different modalities.


\section{Experiments}
This section introduces the experimental setup of this study, including datasets, baseline models, evaluation metrics, and implementation details. Through these settings, we conducted a comprehensive evaluation of the proposed model and compared it with existing state-of-the-art methods to validate its effectiveness.

\subsection{Datasets}
As shown in Table \ref{Table1}, this experiment employs two widely used multimodal dialogue emotion recognition datasets: IEMOCAP and MELD. These datasets contain multimodal features such as text, audio, and vision, making them suitable for evaluating the model's performance in multimodal fusion and dialogue context modeling.

\begin{table*}[h]

\centering
\scriptsize 
\renewcommand\tabcolsep{3pt} 
\renewcommand\arraystretch{1.1} 
\caption{Statistical information of the IEMOCAP and MELD datasets, including the number of conversations, utterances, speakers, total emotion classes, and the number of utterances for each emotion class. The values are reported as Train+Validation / Test splits.}
\label{Table1}
\begin{tabular}{c|c|c|c|c|c|c|c|c|c|c|c|c}
\hline
\textbf{Dataset} & \textbf{Convs} & \textbf{Utterances} & \textbf{Speakers} & \textbf{Classes} & \textbf{Neutral} & \makecell{\textbf{Happy}\\\textbf{Joy}} & \textbf{Sadness} & \textbf{Angry} & \makecell{\textbf{Excited}\\\textbf{Surprise}} & \textbf{Frustrated} & \textbf{Disgust} & \textbf{Fear} \\
\hline \hline
IEMOCAP & 120 / 31 & 5810 / 1623 & 2 & 6 & 1708 / 490 & 1636 / 456 & 1084 / 266 & 1103 / 290 & 1041 / 265 & 1848 / 324 & - & - \\
MELD    & 1153 / 280 & 11098 / 2610 & 9 & 7 & 5180 / 1256 & 1940 / 368 & 794 / 208 & 1243 / 364 & 1205 / 431 & - & 293 / 68 & 268 / 50 \\
\hline
\end{tabular}
\end{table*}

\textbf{IEMOCAP Dataset} \cite{busso2008iemocap}: This dataset was collected by the University of Southern California, consisting of 10 dialogue sessions with a total of approximately 12 hours of video recordings. Each dialogue involves two speakers and is annotated with 6 emotion categories: neutral, happy, sad, angry, excited, and frustrated. The dataset comprises a total of 5531 utterances. We follow the standard partitioning method, using the first 8 sessions as the training set and the last 2 as the test set. The multimodal nature of this dataset makes it a benchmark for evaluating emotion recognition models.

\textbf{MELD Dataset} \cite{poria2018meld}: This dataset is based on dialogue segments from the TV show "Friends," containing approximately 1400 dialogues with 13708 utterances. It is annotated with 7 emotion categories: neutral, surprise, fear, sad, joy, disgust, and angry. We use the official split: approximately 1000 dialogues for the training set, 100 for the validation set, and 300 for the test set. The larger number of speakers in this dataset (up to 9) increases the complexity of dialogue context modeling.

\subsection{Baselines}
To validate the superiority of the proposed model, we selected the following baseline models for comparison. These models cover representative methods based on RNN, GCN, and multimodal fusion. We reproduced them on the same datasets or used publicly available code for experiments.

\textbf{DialogueRNN} \cite{majumder2019dialoguernn}: An RNN-based model that utilizes attention mechanisms to capture emotional dynamics in dialogues.

\textbf{DialogueGCN} \cite{ghosal2019dialoguegcn}: A graph convolutional network-based model that models dialogues as graph structures to capture dependencies between utterances.

\textbf{MMGCN} \cite{hu2021mmgcn}: A multimodal graph attention network that fuses text, audio, and visual modalities through graph attention mechanisms for emotion recognition.

\textbf{MM-DFN} \cite{hu2022mm}: A multimodal dynamic fusion network that uses dynamic attention to fuse multimodal features.

\textbf{COGMEN} \cite{joshi2022cogmen}: A contextualized graph neural network for multimodal emotion recognition, combining GNN and multimodal fusion.

\textbf{MultiEMO} \cite{shi2023multiemo}: An attention-based correlation-aware multimodal fusion framework that emphasizes inter-modal correlations.

\textbf{SDT} \cite{ma2024sdt}: A speaker-dependent Transformer model that uses Transformers to capture speaker-specific patterns.

\textbf{GraphCFC} \cite{li2023graphcfc}: A directed graph-based cross-modal fusion network for multimodal emotion analysis.

\textbf{RL-EMO} \cite{zhang2024rl}: A reinforcement learning-enhanced emotion recognition model that optimizes emotion prediction via RL.

\textbf{DEDNet} \cite{wang2024dynamic}: A dual encoder-decoder network for handling multimodal dialogue emotions.

\textbf{DER-GCN \cite{ai2024gcn}:} The model constructs a weighted multi-relational graph to capture diverse dependencies in dialogues and integrates multimodal features via a self-supervised masked graph autoencoder and a multi-information Transformer.
   
\textbf{MERC-PLTAF \cite{wu2025multi}:} This work adopts fine-grained feature extraction and cross-modal fusion strategies to jointly model multimodal emotional information across dialogues.

\subsection{Evaluation Metrics}

We employ the weighted F1 score and weighted accuracy as the primary evaluation metrics to comprehensively assess model performance. All metrics are computed on the test set using the Scikit-learn library.

\subsection{Implementation Details}
The model is implemented using the PyTorch framework. For the text modality, features are extracted using RoBERTa with a dimension of 1024. For the visual modality, DenseNet is employed to extract features with a dimension of 342; for the audio modality, OpenSmile is used to extract features with a dimension of 1582 for the IEMOCAP dataset or 300 for the MELD dataset. These initial features are projected to a unified hidden dimension of 512 through a linear layer for subsequent processing. To ensure training stability and model performance, the batch size is set to 16, the learning rate is 0.000068, the Adam optimizer is used with a weight decay of 0.00005, and training is conducted for 100 epochs. The modality dropout mechanism has a base probability \( q_{\text{base}} = 0.3 \) and an execution probability \( p_{\text{exe}} = 0.5 \).

\begin{table*}[t]
   \centering
   \caption{The table evaluates the performance of all models on the IEMOCAP (six emotion categories) datasets using F1 scores, while presenting their overall performance across three datasets with weighted accuracy (wa-ACC) and weighted F1 score (wa-F1) as metrics, where the best results are bolded and the second-best are underlined.}
   \resizebox{\textwidth}{!}{
   \begin{tabular}{l|cc|cc|cc|cc|cc|cc|cc}
   \hline
   \multirow{3}{*}{Models} & \multicolumn{12}{c|}{IEMOCAP} & \multicolumn{2}{c}{} \\
   \cline{2-15}
   & \multicolumn{2}{c|}{happy} & \multicolumn{2}{c|}{sad} & \multicolumn{2}{c|}{neutral} & \multicolumn{2}{c|}{angry} & \multicolumn{2}{c|}{excited} & \multicolumn{2}{c|}{frustrated} & \multirow{2}{*}{wa-ACC} & \multirow{2}{*}{wa-F1} \\
   \cline{2-13}
   & ACC & F1 & ACC & F1 & ACC & F1 & ACC & F1 & ACC & F1 & ACC & F1 \\
   \hline
   DialogueRNN & 25.00 & 34.95 & 82.86 & \underline{84.58} & 54.43 & 57.66 & 61.76 & 64.42 & \textbf{90.97} & 76.30 & 62.20 & 59.55 & 65.43 & 64.29 \\
   DialogueGCN & 64.29 & 29.03 & 80.86 & 64.37 & 43.14 & 50.96 & 68.49 & 63.29 & 71.85 & 68.19 & 57.68 & 62.41 & 62.07 & 58.19 \\
   MMGCN & 32.64 & 39.66 & 72.65 & 76.89 & 65.10 & 62.81 & 73.53 & 71.43
   & 77.93 & 75.40 & 65.09 & 63.43 & 66.62 & 66.25 \\
   MM-DFN & 44.44 & 44.44 & 77.55 & 80.00 & 71.35 & 66.99 & \underline{75.88} & 70.88 & 74.25 & 76.42 & 58.27 & 61.67 & 67.84 & 67.85 \\
   TS-GCL & \underline{71.20} & \underline{70.00} & 81.30 & 81.70 & 67.40 & 64.20 & 60.50 & 61.40 & 74.60 & 76.50 & 62.00 & 64.00 & 70.30 & 70.20 \\
   MultiEMO & 53.80 & 56.29 & 83.33 & 83.50 & \textbf{75.60} & 70.11 & 68.29 & 67.07 & 79.70 & 75.79 & 64.82 & 70.35 & 72.29 & 71.69 \\
   SDT & 55.06 & 57.62 & 80.58 & 80.08 & 65.73 & 69.14 & 67.88 & 66.87 & 82.50 & 73.47 & 66.58 & 67.53 & 70.54 & 69.95 \\
   GraphCFC & 41.52 & 45.08 & \textbf{87.12} & \textbf{84.84} & 65.19 & 63.27 & 68.31 & 70.82 & 77.16 & 75.85 & 62.86 & 63.19 & 68.39 & 68.02 \\
   RL-EMO & 40.28 & 47.15 & 79.18 & 81.17 & 69.79 & 66.67 & 74.12 & 64.28 & 78.60 & 76.18 & 58.01 & 61.82 & 69.16 & 68.20 \\
   DEDNet & 56.32 & 64.07 & 81.15 & 80.98 & 73.92 & \underline{74.97} & 67.37 & 71.11 & 84.38 & \underline{77.84} & 72.99 & 69.68 & \underline{74.47} & \underline{73.79} \\

DER-GCN & 60.70 & 58.80 & 75.90 & 79.80 & 66.50 & 61.50 & 71.30 & \underline{72.10} & 71.10 & 73.30 & 66.10 & 67.80 & 69.70 & 69.40 \\
MERC-PLTAF & \textbf{75.60} & \textbf{75.30} & 74.20 & 80.00 & 71.10 & 71.40 & \textbf{75.90} & \textbf{74.70} & 59.70 & 54.70 & \textbf{73.50} & \textbf{72.20} & 72.70 & 71.40 \\

   \hline
   AMB-DSGDN(Ours)& 60.23 & 66.25 & \underline{84.58} & 81.36 & \underline{74.14} & \textbf{76.20} & 70.86 & 71.88 & \underline{85.87} & \textbf{81.34} & \underline{73.24} & \underline{72.17} & \textbf{76.09} & \textbf{75.64} \\
   \hline
   \end{tabular}
   }
   \label{Table2}
   \end{table*}

\begin{table*}[t]
\caption{The table evaluates the performance of all models on the MELD (seven emotion categories) datasets using F1 scores, while presenting their overall performance across three datasets with wa-ACC and wa-F1 as metrics, where the best results are bolded and the second-best are underlined.}
\resizebox{\textwidth}{!}{
\begin{tabular}{l|cc|cc|cc|cc|cc|cc|cc|cc}
\hline
\multirow{3}{*}{Models} & \multicolumn{14}{c|}{MELD} & \multicolumn{2}{c}{} \\
\cline{2-17}
& \multicolumn{2}{c|}{neutral} & \multicolumn{2}{c|}{surprise} & \multicolumn{2}{c|}{fear} & \multicolumn{2}{c|}{sadness} & \multicolumn{2}{c|}{joy} & \multicolumn{2}{c|}{disgust} & \multicolumn{2}{c|}{anger} & \multirow{2}{*}{wa-ACC} & \multirow{2}{*}{wa-F1} \\
\cline{2-15}
& ACC & F1 & ACC & F1 & ACC & F1 & ACC & F1 & ACC & F1 & ACC & F1 & ACC & F1 \\
\hline
DialogueRNN & 82.17 & 76.56 & 46.62 & 47.64 & 0.00 & 0.00 & 21.15 & 24.65 & 49.50 & 51.49 & 0.00 & 0.00 & 48.41 & 46.01 & 60.27 & 57.95 \\
MMGCN & \underline{84.32} & 76.96 & 47.33 & 49.63 & 2.00 & 3.64 & 14.90 & 20.39 & 56.97 & 53.76 & 1.47 & 2.82 & 42.61 & 45.23 & 61.34 & 58.41 \\
MM-DFN & 79.06 & 75.80 & 53.02 & 50.42 & 0.00 & 0.00 & 17.79 & 23.72 & 59.20 & 55.48 & 0.00 & 0.00 & 50.43 & 48.27 & 60.96 & 58.72 \\
TS-GCL & 78.10 & \textbf{80.60} & 56.70 & 56.40 & 6.80 & 5.20 & 42.30 & \underline{43.70} & \textbf{68.30} & \underline{66.30} & 2.30 & 2.60 & 43.80 & 48.50 & 64.40 & 64.10 \\
MultiEMO & 76.44 & 79.42 & 55.92 & 58.12 & \underline{25.71} & 21.18 & 51.05 & 41.60 & 62.23 & 63.07 & \textbf{45.71} & \underline{31.07} & \underline{55.28} & 53.37 & 65.45 & 65.77 \\
SDT & 75.99 & 79.65 & \textbf{59.21} & 58.78 & 24.44 & \underline{23.16} & \textbf{59.22} & 39.23 & 64.40 & 62.76 & 40.00 & \textbf{31.86} & 52.27 & \underline{54.44} & 66.00 & \underline{65.92} \\
GraphCFC & 71.26 & 75.17 & 46.18 & 45.68 & 9.09 & 3.28 & 30.43 & 11.42 & 50.26 & 49.49 & 0.00 & 0.00 & 35.88 & 41.93 & 54.34 & 55.20 \\
RL-EMO & \textbf{85.59} & 79.57 & \underline{58.72} & 59.03 & 14.00 & 16.09 & 18.27 & 27.64 & 60.45 & 63.53 & 16.18 & 20.18 & \textbf{55.36} & 52.84 & 65.63 & 63.47 \\
DEDNet & 76.51 & 79.97 & 56.77 & 58.90 & 21.88 & 17.07 & 50.38 & 39.30 & \underline{64.97} & 62.63 & 40.54 & 28.57 & 53.65 & \textbf{54.49} & 65.52 & 65.88 \\
MERC-PLTAF & 82.90 & \underline{80.50} & 59.10 & \underline{59.10} & 24.00 & \textbf{26.90} & 55.30 & \textbf{46.40} & 63.40 & \textbf{77.10} & 17.60 & 26.10 & 50.30 & 53.90 & \textbf{68.00} & 52.80\\

\hline
AMB-DSGDN(Ours) & 75.16 & 80.00 & 58.48 & \textbf{59.30} & \textbf{30.00} & 17.14 & \underline{58.18} & 40.25 & 64.75 & 64.59 & \underline{44.44} & 25.26 & 54.93 & 54.12 & \underline{66.07} & \textbf{66.18} \\
\hline
\end{tabular}
}
\label{Table3}
\end{table*}

\section{Results and analysis}

This section systematically evaluates the proposed AMB-DSGDN model from multiple perspectives, including overall performance, ablation studies, parameter sensitivity, robustness, and computational complexity, to comprehensively analyze its effectiveness and practicality.

\subsection{Main Results}

Tables \ref{Table2} and \ref{Table3} show the performance of the proposed model AMB-DSGDN on the IEMOCAP and MELD datasets, compared with various baseline models. To ensure experimental fairness, all models use the same preprocessing and training settings.

On the IEMOCAP dataset, AMB-DSGDN achieved a wa-ACC of 76.09\% and a wa-F1 of 75.64\%, improving by 1.62\% and 1.85\% respectively compared to the second-best model DEDNet. Especially on anger, excitement, and frustration emotions, the model shows significant advantages, benefiting from the differential attention graph convolutional network's ability in fine-grained semantic relation modeling, as well as the adaptive modality dropout strategy's effective alleviation of modality imbalance, enabling the model to better capture key emotional information in dialogues.

On the MELD dataset, AMB-DSGDN's weighted accuracy is 66.07\%, and weighted F1 score is 66.18\%, with limited improvements compared to the second-best model. We believe there are multiple factors contributing to this. First, as shown in Table 1, MELD's class distribution is highly imbalanced, with sparse samples for several emotions (such as disgust, fear), limiting the model's learning and generalization on low-frequency categories; Second, MELD is multi-speaker dialogue, where audio and visual signals are more susceptible to interference from speaker switching, overlapping speech, facial occlusion, etc., thereby reducing unimodal quality and increasing modality fusion difficulty; Nevertheless, the model still shows obvious advantages on categories like surprise, demonstrating its adaptability in multi-speaker environments.

Overall, AMB-DSGDN shows significant advantages on IEMOCAP due to higher modality quality and clearer emotional associations; while on MELD, influenced by class distribution imbalance and modality quality fluctuations caused by multi-speakers, overall improvements are limited, but it still has advantages on categories like surprise, reflecting the model's robustness and adaptability.

\subsection{Ablation Study}
\label{sec:sensitivity_warmup} 

To thoroughly validate the effectiveness of key components in the model, we conduct a series of ablation experiments, focusing on factors such as modality combinations, window sizes, differential attention mechanisms, and adaptive modality dropout strategies. By progressively removing or adjusting these modules and retraining/evaluating on the IEMOCAP and MELD datasets, we quantify the contributions of each component and their synergistic effects.

\begin{table}[h]
\centering
\setlength{\tabcolsep}{2.8pt}  
\renewcommand{\arraystretch}{1.1}
\caption{Comparison of performance of different modal combinations on IEMOCAP and MELD datasets.}
\begin{tabular}{>{\centering\arraybackslash}p{2.4cm}|cc|cc}
\hline
\multirow{2}{*}{\textbf{Modality Setting}}
& \multicolumn{2}{c|}{\textbf{IEMOCAP}}
& \multicolumn{2}{c}{\textbf{MELD}} \\ \cline{2-3} \cline{4-5}
& \textbf{wa-ACC} & \textbf{wa-F1}
& \textbf{wa-ACC} & \textbf{wa-F1} \\ \hline
A          & 62.60 & 60.70 & 34.70 & 37.70 \\
T          & 69.02 & 68.63 & 65.61 & 65.78 \\
V          & 39.80 & 35.60 & 23.16 & 31.27 \\ \hline
A+T        & 73.43 & 72.87 & 65.49 & 65.65 \\
V+T        & 69.71 & 69.16 & 65.77 & 65.86 \\ \hline
\textbf{A+V+T}
& \textbf{76.09} & \textbf{75.64}
& \textbf{66.07} & \textbf{66.18} \\ \hline
\end{tabular}
\label{Table4}
\end{table}
\textbf{Multimodal Combinations:} In Table \ref{Table4}, we examine the impact of different modality combinations on model performance to verify the effectiveness of multimodal fusion. Specifically, we evaluate performance under unimodal, bimodal, and full-modal settings and compare across the two datasets. The results show that in unimodal configurations, the text modality generally achieves the best performance, reflecting its advantage in capturing dialogue semantics and contextual information, followed by the audio modality, likely due to its sensitivity to speech tone and rhythm. In contrast, the visual modality performs relatively weakly, possibly because visual cues (such as facial expressions) are limited by perspectives and dynamic changes in dialogue scenes, leading to instability in information extraction. Furthermore, in bimodal settings, the combination of audio and text significantly outperforms the visual-text combination, indicating that the audio modality better complements textual information by providing additional acoustic cues to enhance emotion recognition accuracy. While the visual-text fusion shows improvements, the gains are limited, suggesting that the visual modality may introduce noise or redundancy in certain dialogue contexts. Ultimately, full-modal integration (i.e., combining audio, visual, and text) achieves the best performance on both datasets, validating the importance of multimodal synergy. By leveraging multiple information sources simultaneously, the model can more comprehensively capture emotional dynamics, alleviating the limitations of unimodal or bimodal approaches and thereby improving overall robustness and generalization.

\begin{figure}[t]
	\includegraphics[width=0.99\linewidth]{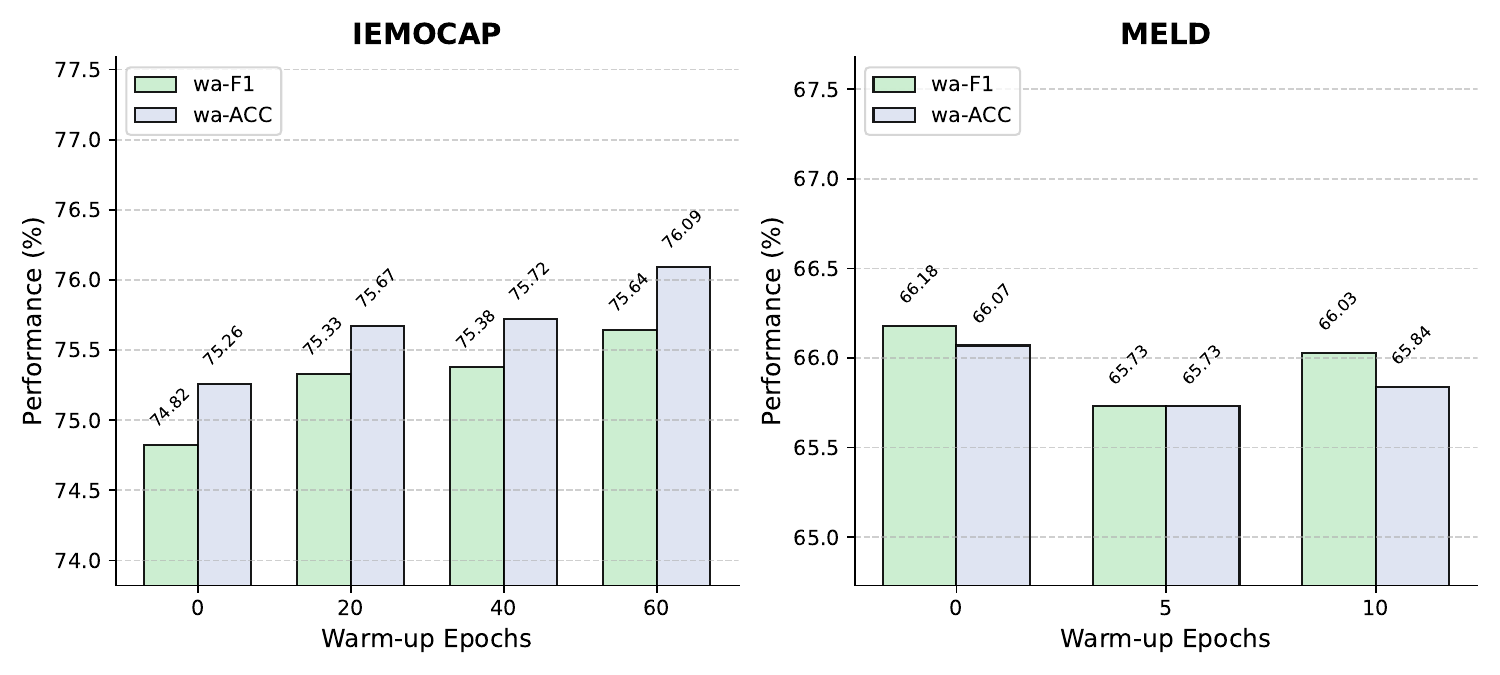}
	\caption{Experimental results of different window sizes on IEMOCAP and MELD datasets. The purple line represents wa-ACC, the light yellow bar represents wa-F1; the left subplot corresponds to IEMOCAP dataset, the right subplot to MELD dataset. Note: Window size determines the semantic association capture range of the graph convolutional network and needs to be adjusted based on dataset characteristics.}
	\label{fig3}
\end{figure}

\textbf{Window Size Settings:} To explore the impact of contextual range on graph convolutional networks in dialogue emotion recognition tasks, this experiment systematically evaluates the role of different window sizes on model performance. Fig. \ref{fig3} shows that on the IEMOCAP dataset, as window size gradually increases, the model's weighted average accuracy and F1 values reach optimal performance under medium window configurations. This indicates that moderate windows can fully cover semantically related utterances and their contextual dependencies, thereby improving the graph structure's modeling ability for emotional relationships. Too small windows limit the capture of long-range dependency information, leading to insufficient semantic information; too large windows may introduce irrelevant features, accumulating noise, thereby reducing discriminative ability in attention weight calculation and feature propagation. On the MELD dataset, optimal performance is concentrated in relatively smaller window intervals. Considering that MELD's dialogues have more frequent inter-speaker interactions and more complex multi-turn structures, smaller windows can focus more on local semantics and multimodal information interactions, while overly large windows easily incorporate irrelevant emotional features across speakers, increasing modality inconsistencies and distribution noise. This shows that different datasets' dialogue structures, speaker distributions, and emotional interaction patterns significantly affect the optimal choice of window size, and in practical applications, this hyperparameter needs to be tuned based on data statistical characteristics.

\begin{table}[t]
\centering
\caption{Ablation Study of Semantic Graph Differential Network on IEMOCAP and MELD Datasets.}
\resizebox{0.48\textwidth}{!}{
\begin{tabular}{l|cc|cc}
\hline
\multirow{2}{*}{Variants} & \multicolumn{2}{c|}{IEMOCAP} & \multicolumn{2}{c}{MELD} \\
\cline{2-5}
& wa-F1 & wa-ACC & wa-F1 & wa-ACC \\
\hline
w/o Rel Subgraph + DiffRGCN & 68.00 & 68.82 & 65.07 & 65.10 \\
w/o DiffRGCN & 73.44 & 73.56 & 65.62 & 64.80 \\
Full Model & \textbf{75.64} & \textbf{76.09} & \textbf{66.07} & \textbf{66.18} \\
\hline
\end{tabular}
}
\label{Table5}
\end{table}

\textbf{Semantic Graph Differential Network:} To verify the role of DiffRGCN and relational subgraph components in the overall model, we designed targeted ablation studies by replacing or removing related modules while keeping the rest of the structure consistent to analyze the independent contributions of each component to performance. It should be noted that the adaptive modality balancing mechanism remains enabled in this group of ablation studies. Specifically, when only DiffRGCN is removed, we replace the differential relational graph convolution with a regular graph convolutional neural network for feature propagation on the constructed relational subgraphs, with the rest of the model architecture unchanged; when both relational subgraphs and DiffRGCN are removed, the model no longer performs graph structure modeling, but directly connects the adaptive modality balancing mechanism and subsequent classification modules after the utterance-level multimodal encoder. The results in Table \ref{Table5} show that simultaneously removing relational subgraphs and DiffRGCN causes significant declines in weighted F1 and accuracy on IEMOCAP and MELD, verifying the importance of semantic relation modeling. Further comparison reveals that replacing DiffRGCN with GCN alone leads to performance degradation but still outperforms the variant without graph structures entirely, indicating that relational subgraphs provide basic semantic associations, while DiffRGCN utilizes differential attention to characterize fine-grained semantic differences between nodes, effectively filtering redundant information and strengthening the representation of multimodal emotional dependencies. The above results fully prove the key value of the semantic graph differential network in multimodal learning and its positive impact on model performance and generalization capabilities.

\begin{figure}[t]
	\includegraphics[width=0.99\linewidth]{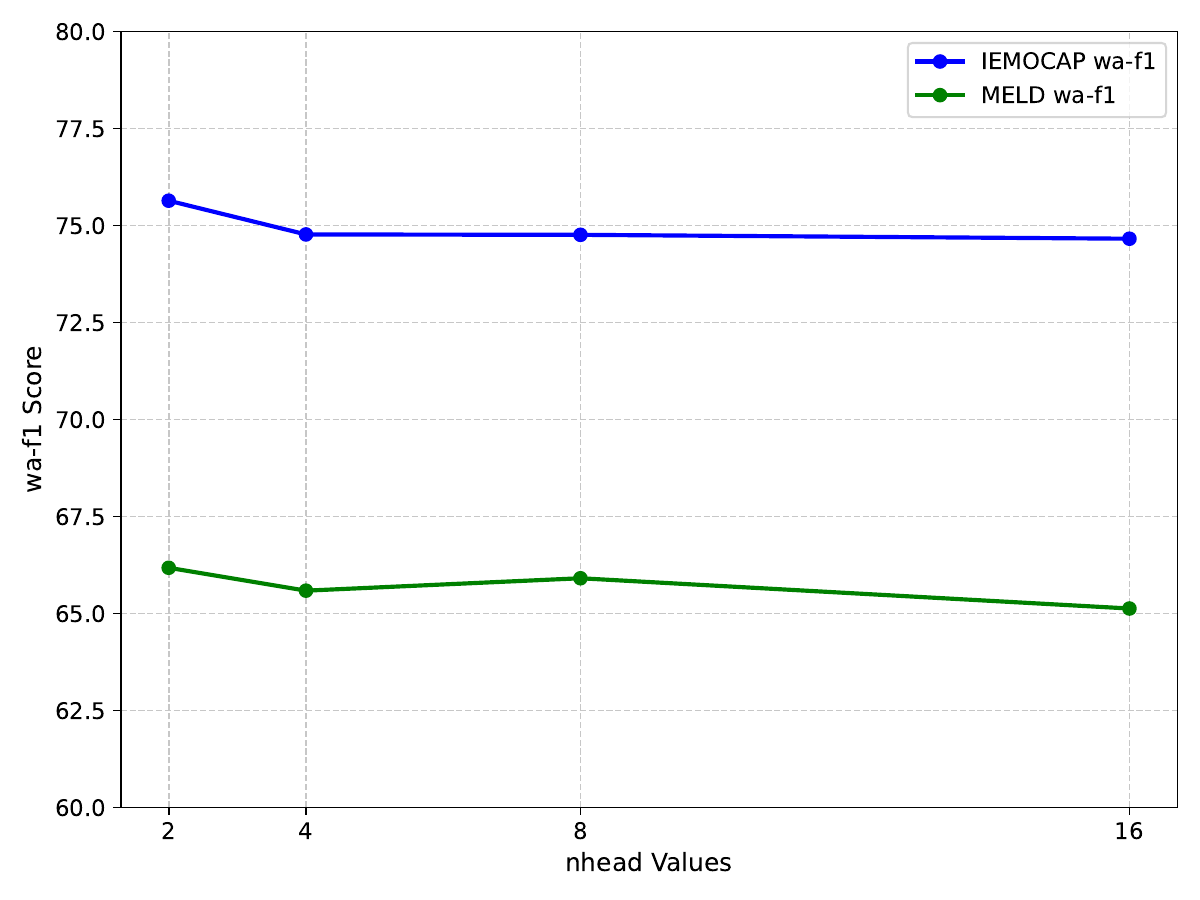}
	\caption{Sensitivity analysis results of attention head numbers on IEMOCAP and MELD datasets. The blue line represents wa-F1 on IEMOCAP dataset, the green line represents wa-F1 on MELD dataset, horizontal axis “nhead Values” indicates number of attention heads, vertical axis “wa-F1 Score” indicates corresponding weighted average F1 score.}
	\label{fig4}
\end{figure}

\textbf{Sensitivity of Attention Heads:} In further sensitivity analysis, we explored the impact of the number of attention heads in the differential attention graph convolutional network on model performance to reveal its role in multimodal feature aggregation. As shown in Fig. \ref{fig4}, on the IEMOCAP dataset, the model performs best under a small number of attention heads. This is because a small number of attention heads can effectively capture semantic diversity while keeping feature representations compact. Too few heads may lead to insufficient multimodal information integration, while too many heads may introduce redundant features or increase training difficulty, thereby reducing model discriminative ability. On the MELD dataset, the model also performs excellently under low head configurations, with performance slightly declining as the number of heads increases. Attributed to the multi-speaker dialogue characteristics of the MELD dataset, the integration difficulty of cross-modal signals is higher in multi-speaker environments, too many attention heads may introduce unnecessary noise and information conflicts, while a streamlined attention mechanism can more concentratedly capture core semantic associations and emotional dynamics, achieving efficient information integration.

\begin{table}[t]
\centering
\caption{Ablation Study of Modality Balancing on IEMOCAP and MELD Datasets.}
\resizebox{0.48\textwidth}{!}{
\begin{tabular}{l|cc|cc}
\hline
\multirow{2}{*}{Variants} & \multicolumn{2}{c|}{IEMOCAP} & \multicolumn{2}{c}{MELD} \\
\cline{2-5}
& wa-F1 & wa-ACC & wa-F1 & wa-ACC \\
\hline
w/o MD + Rel Subgraph + DiffRGCN & 67.83 & 68.78 & 64.87 & 64.80 \\
w/o MD + DiffRGCN & 72.99 & 73.56 & 65.56& 65.71 \\
w/o MD & 74.77 & 75.29 & 66.00 & 66.05 \\
Full Model & \textbf{75.64} & \textbf{76.09} & \textbf{66.07} & \textbf{66.18} \\
\hline
\end{tabular}
}
\label{Table6}
\end{table}

\textbf{Adaptive Modality Balancing:} As shown in Table \ref{Table6}, we conducted ablation analysis on the synergistic effects of the adaptive modality dropout (MD) strategy and semantic graph components (relational subgraphs and DiffRGCN) by progressively removing core modules. In this group of experiments, the adaptive modality balancing mechanism is disabled by default, with the rest of the network structure and experimental settings consistent with the semantic graph differential network subsection. Specifically, when simultaneously removing the modality dropout mechanism, relational subgraphs, and DiffRGCN, the model no longer performs modality regulation or graph structure modeling, but directly inputs utterance-level multimodal representations into the classification module; when removing the modality dropout mechanism and DiffRGCN, the model only retains relational subgraphs for feature aggregation; when only removing the modality dropout mechanism, the model retains the complete semantic graph structure, but each modality participates in fusion with fixed weights. The experimental results show that simultaneously removing all three leads to the maximum decline in weighted F1 and accuracy on IEMOCAP and MELD, indicating significant complementarity between modality regulation and semantic relation modeling. Further comparison reveals that the model without the modality dropout mechanism still experiences performance decline in multimodal feature fusion, showing that this mechanism can dynamically adjust each modality's participation, suppressing excessive influence from dominant modalities, thereby improving emotion recognition accuracy under noise interference. In summary, the adaptive modality dropout strategy and semantic graph components have significant complementary effects in multimodal dialogue emotion recognition, synergistically enhancing the model's ability to capture emotional dynamic dependencies and robustness to noise interference.

\begin{figure}[t]
	\includegraphics[width=0.99\linewidth]{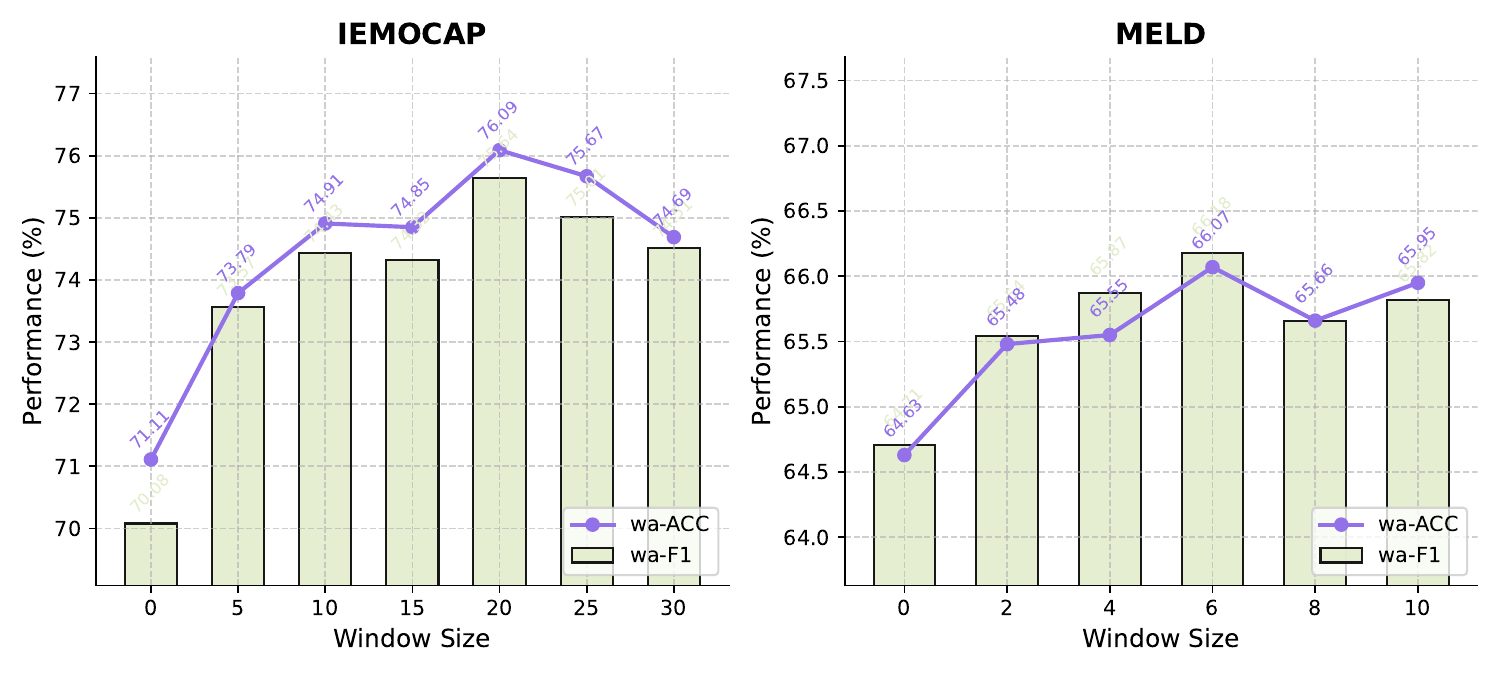}
	\caption{Performance dynamics in the warm-up phase. The horizontal axis represents the warm-up period (i.e., the number of training rounds before the modality dropout mechanism is gradually activated), the vertical axis is model performance metrics; different color curves correspond to weighted average F1 scores and accuracies on IEMOCAP and MELD datasets respectively, showing the impact of different warm-up periods on model performance.}
	\label{fig5}
\end{figure}

\textbf{Performance Dynamics During Warm-up Period:} To verify the effectiveness of the adaptive modality dropout strategy, we examined the impact of the warm-up period (i.e., the number of training rounds before the modality dropout mechanism is gradually activated) on model performance to reveal its role in model stability and optimization effects in the early training stage. As shown in Fig. \ref{fig5}, on the IEMOCAP dataset, as the warm-up period extends, the model's performance first rises rapidly, reaching a peak at about 60 training rounds, then remains stable. This indicates that moderate warm-up can help the model gradually adapt to modality imbalance issues, reducing noise interference in early training, thereby improving the capture accuracy of emotional dynamic features. If the warm-up period is too short, the model may enter the modality dropout phase prematurely, leading to insufficient learning of internal representations and cross-modal relationships in each modality's features, resulting in performance fluctuations or unstable convergence. For the MELD dataset, the model's initial performance is higher without warm-up, with performance slowly declining as warm-up rounds increase, and optimal performance appears in shorter warm-up period configurations. Due to the inherent complexity of multi-speaker dialogues, the integration difficulty of cross-modal signals increases accordingly, overly long warm-up periods may cause the model to overly rely on partial modality features without sufficiently focusing on core emotional cues, thereby introducing redundant information interference. Conversely, shorter warm-up periods can quickly activate the modality dropout mechanism, enabling the model to more effectively identify and weaken redundant features, thereby efficiently integrating cross-modal information, improving training convergence speed, and stabilizing performance.

\begin{table}[h]
\centering
\caption{Sensitivity analysis results of modal discard probability parameter.}
\begin{tabular}{c|c|c|c}
\hline
\multicolumn{4}{c}{\textbf{IEMOCAP}} \\
\hline
$q_{base}$ & $p_{exe}=0.3$ & $p_{exe}=0.5$ & $p_{exe}=0.7$ \\
\hline
0.1 & 74.88 & 75.62 & 75.2 \\
0.3 & 75.37 & \textbf{75.64} & 75.06 \\
0.5 & 75.07 & 74.96 & 75.12 \\
\hline
\hline
\multicolumn{4}{c}{\textbf{MELD}} \\
\hline
$q_{base}$ & $p_{exe}=0.1$ & $p_{exe}=0.2$ & $p_{exe}=0.3$ \\
\hline
0.1 & 65.91 & 65.8 & 65.78 \\
0.2 & 65.80 & \textbf{66.18}& 65.81 \\
0.3 & 65.87 & 65.87 & 65.72 \\
\hline
\end{tabular}
\label{Table7}
\end{table}

\textbf{Dropout Probability Settings:} In the parameter sensitivity analysis of the adaptive modality dropout strategy, we systematically explored the impact of different combinations of base dropout probability q\_base and execution probability p\_exe on model performance through grid search on two datasets to evaluate its optimization effects. As shown in Table \ref{Table7}, on the IEMOCAP dataset, model performance gradually improves under lower base dropout probability configurations, reaching a peak under medium base dropout probability (e.g., 0.3) and medium execution probability (e.g., 0.5) combinations, then stabilizing. This indicates that moderate dropout can balance noise suppression and key information retention, too low dropout may lead to noise features interfering with fusion results, while too high dropout disrupts semantic representations and increases training instability. Similarly, on the MELD dataset, the model has a higher starting performance under lower base dropout probabilities, with optimal configurations concentrated in relatively lower base dropout probabilities (e.g., 0.2) and medium execution probability intervals, closely related to the dataset's multi-speaker, rich dialogue content, and diverse emotion characteristics. Conservative dropout strategies can effectively retain core information from each modality, reducing interference caused by signal differences between different speakers, while balancing noise suppression, making the model more robust in complex and diverse dialogue environments.

\begin{figure*}[t]
	\includegraphics[width=0.99\linewidth]{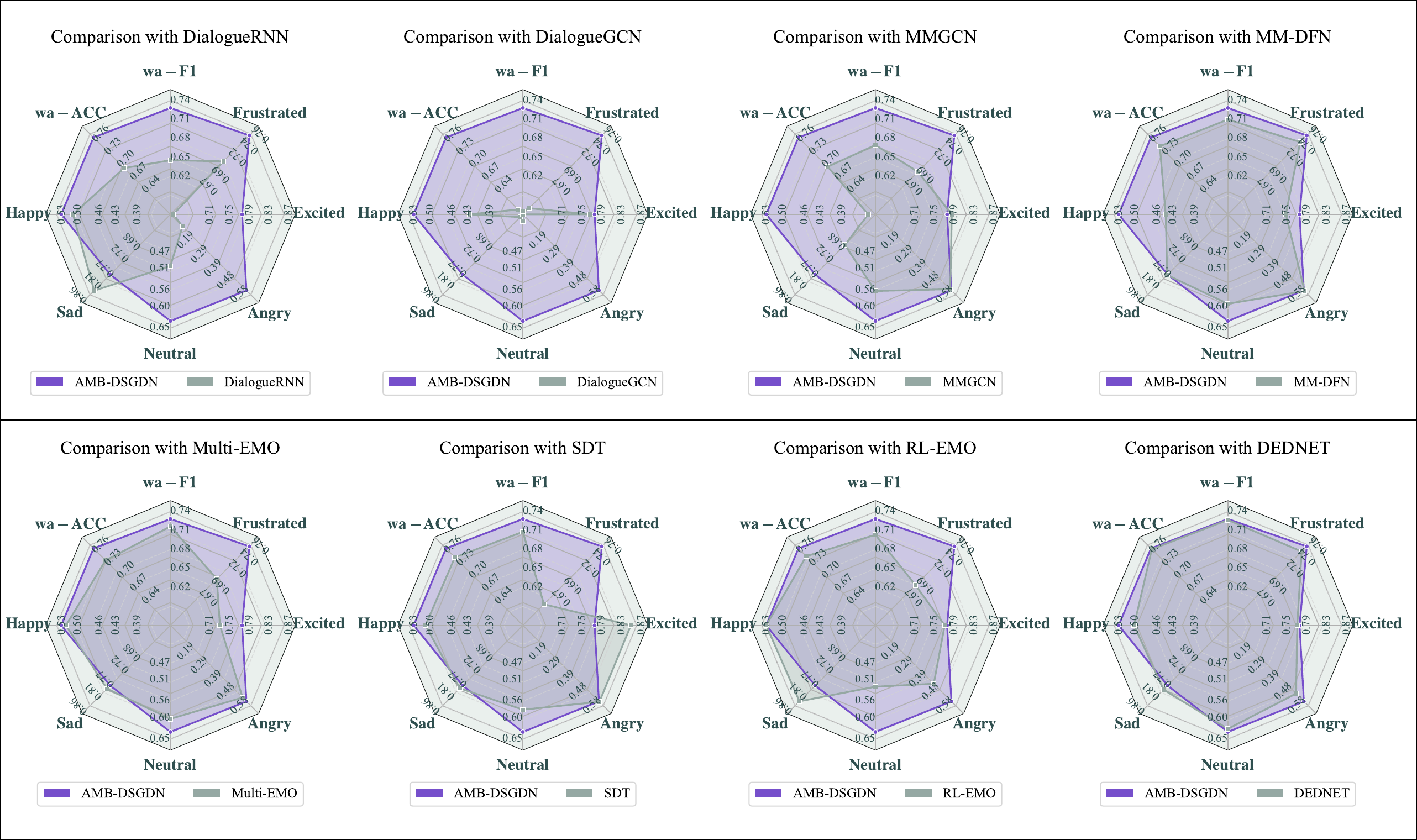}
	\caption{Performance comparison of the proposed method with other baseline methods on the long sequence dialogue subset of the IEMOCAP dataset, intuitively showing the performance advantages of the proposed method in experiments.}
	\label{fig6}
\end{figure*}

\subsection{Handling Long-Sequence Dialogues}

This section conducts a dedicated evaluation of the model's capability in long-sequence dialogue scenarios, focusing on its ability to capture dynamic contextual dependencies. The experiments are based on a long-dialogue subset of the IEMOCAP dataset, selecting dialogue samples with 20 to 50 utterances to simulate the complex and evolving temporal dependencies found in real interactive scenarios. Model performance is assessed using weighted F1 score and accuracy, and compared against multiple baseline methods.

The results shown in Fig. \ref{fig6} indicate that the proposed method significantly outperforms the baseline models on the long-sequence dialogue subset, demonstrating its effectiveness in modeling long-range contextual dependencies. Compared with the baselines, our model maintains stable performance even on longer dialogue sequences, without noticeable degradation, indicating stronger capability in capturing long-term dependencies.

Further analysis reveals that this advantage primarily arises from two design aspects. First, the dynamic cross-modal contribution balancing mechanism adjusts the participation of different modalities in real time according to the dialogue progression, preventing a single modality from dominating and causing information bias over long sequences. Second, the differential relation graph modeling explicitly captures fine-grained differences between nodes, effectively mitigating the common issue of context forgetting in long-sequence tasks. In the middle stages of dialogue, baseline methods often exhibit performance fluctuations due to accumulated noise and modality shifts, whereas the proposed method maintains stable emotion recognition by adaptively adjusting modality weights.

In the later stages of dialogue, our method makes more effective use of accumulated contextual information, suppressing the influence of early noise on current predictions and reducing error propagation. This indicates that the model is capable not only of capturing local emotional changes but also of modeling global emotional evolution trends effectively.

From the perspective of emotional dependency evolution, emotions in long dialogues exhibit clear non-linear and phased characteristics. In the early stages, the model primarily relies on local modality features for judgment. As the dialogue progresses, emotional interactions between speakers gradually intensify, and the differential mechanism captures changes in the positive and negative subspaces. In the later stages, the model strengthens critical contextual information through dynamic attention, effectively integrating information across the entire sequence. This process ensures a comprehensive modeling of the emotional trajectory in the dialogue, further validating the robustness and adaptability of the proposed method in long-sequence dialogue tasks.

\subsection{Computational Overhead Increment Analysis} To comprehensively evaluate the computational impact of the proposed differential graph convolution network and the adaptive modality balancing module, we conduct a comparative analysis of the full model on the IEMOCAP and MELD datasets in terms of inference time, batch processing time, throughput, and floating-point operations per second (FLOPs), as reported in Table \ref{Table8}.
\begin{table*}[t]
\centering
\small
\renewcommand{\arraystretch}{1.45}
\setlength{\tabcolsep}{4pt}
\caption{Computational overhead analysis of DiffRGCN compared to the standard GAT, and of the Adaptive Modality Balancing module.}
\label{Table8}
\begin{tabular}{
@{}c@{\hspace{3pt}}c@{\hspace{6pt}}
cc cc cc cc}
\toprule
\multirow{2}{*}{\textbf{Module}} 
& \multirow{2}{*}{\textbf{Dataset}} 
& \multicolumn{2}{c}{\textbf{Inference Time (ms)}} 
& \multicolumn{2}{c}{\textbf{Throughput (samples/s)}} 
& \multicolumn{2}{c}{\textbf{Batch Time (ms)}} 
& \multicolumn{2}{c}{\textbf{FLOPs/s (T)}} \\
\cmidrule(lr){3-4} \cmidrule(lr){5-6} \cmidrule(lr){7-8} \cmidrule(lr){9-10}
& 
& \textbf{Baseline} & \textbf{$\Delta$ (Change)} 
& \textbf{Baseline} & \textbf{$\Delta$ (Change)}
& \textbf{Baseline} & \textbf{$\Delta$ (Change)}
& \textbf{Baseline} & \textbf{$\Delta$ (Change)} \\
\midrule
\multirow{2}{*}{\makecell[c]{DiffRGCN}}
& IEMOCAP
& 0.0769
& \makecell[c]{+0.0090\\(+11.70\%)}
& 13169.51
& \makecell[c]{$-$1693.29\\($-$12.86\%)}
& 62.14
& \makecell[c]{+10.7027\\(+17.22\%)}
& 0.0805
& \makecell[c]{$-$0.0157\\($-$19.50\%)} \\
& MELD
& 0.4271
& \makecell[c]{+0.1026\\(+24.03\%)}
& 2384.02
& \makecell[c]{$-$404.02\\($-$16.95\%)}
& 60.33
& \makecell[c]{+11.4914\\(+19.05\%)}
& 0.0500
& \makecell[c]{$-$0.00236\\($-$4.72\%)} \\
\midrule
\multirow{2}{*}{\makecell[c]{Adaptive Modality\\Balancing}}
& IEMOCAP
& 0.0769
& \makecell[c]{+0.0165\\(+21.45\%)}
& 13169.51
& \makecell[c]{$-$980.73\\($-$7.45\%)}
& 62.14
& \makecell[c]{+6.52\\(+10.50\%)}
& 0.0805
& \makecell[c]{$-$0.0082\\($-$10.19\%)} \\
& MELD
& 0.4271
& \makecell[c]{+0.0190\\(+4.45\%)}
& 2384.02
& \makecell[c]{$-$137.90\\($-$5.79\%)}
& 60.33
& \makecell[c]{+2.1635\\(+3.59\%)}
& 0.0500
& \makecell[c]{$-$0.00384\\($-$7.68\%)} \\
\bottomrule
\end{tabular}
\end{table*}

For the differential graph convolution network, compared with the conventional graph attention network, the additional computational overhead mainly arises from the introduction of the differential attention mechanism and relation embedding computation. By modeling the differences among multiple attention distributions, this mechanism effectively highlights context-relevant dependencies, thereby enhancing the modeling of intra-speaker emotional continuity and inter-speaker emotional interactions. As each node is required to compute multiple sets of attention weights and perform differential operations, the computational complexity increases with the graph size and the number of edges, resulting in a moderate increase in inference time and batch processing time, accompanied by a slight decrease in throughput. Notably, this overhead is primarily confined to attention computations within local subgraphs and does not introduce global high-order operations, leading to a relatively mild growth in complexity as the dialogue length increases. In long-sequence dialogue scenarios, DiffRGCN can still effectively limit the computational scope through window-based subgraph modeling, thereby maintaining good scalability.

For the adaptive modality balancing module, its core objective is to dynamically regulate the relative contributions of different modalities during multimodal fusion via an adaptive modality dropout strategy, alleviating the dominance of a single modality in the decision-making process. The module mainly consists of lightweight operations, including probability estimation, stochastic sampling, and feature scaling, without introducing additional complex network structures or large-scale matrix operations, and thus imposes a limited computational burden on the overall model. In long-sequence scenarios, its computational complexity scales approximately linearly with the sequence length, enabling the model to maintain stable computational efficiency as dialogue length increases. By dynamically adjusting modality contributions while preserving computational efficiency, this module further enhances the model’s generalization ability and robustness in complex conversational environments.

\begin{figure}[!t]
	\includegraphics[width=0.99\linewidth]{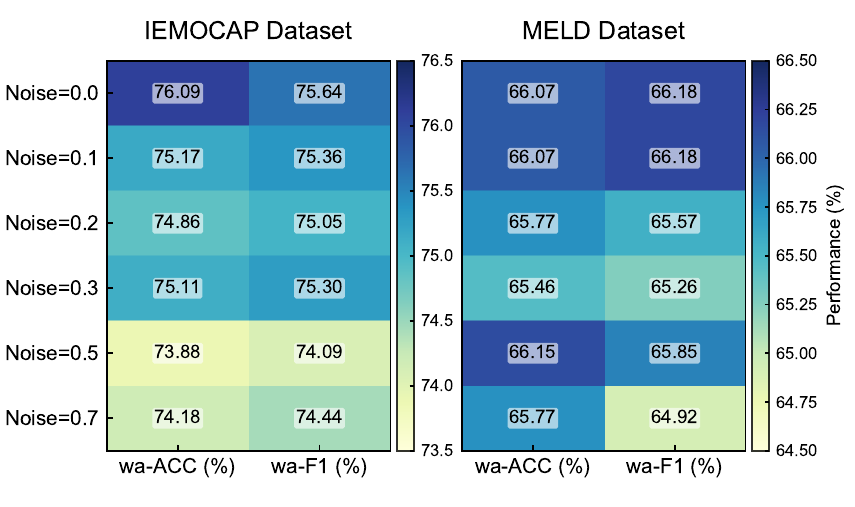}
	\caption{Full-modality noise robustness test results of the model on the IEMOCAP and MELD datasets. Each cell in the figure shows the model’s wa-ACC and wa-F1 under different noise intensities. The noise intensity ranges from 0 to 0.7, representing the standard deviation multiples of Gaussian noise added to the textual, visual, and audio features during testing. The color encoding indicates performance levels, with darker colors representing higher performance; the color bar on the right provides a reference for the corresponding percentage values. This heatmap intuitively illustrates the model’s stability and performance variations under different noise conditions.}
	\label{fig7}
\end{figure}

\subsection{Robustness Analysis of Multimodal Noise:} To evaluate the model’s stability and generalization ability under complex noisy conditions, we conducted full-modality noise interference experiments on the IEMOCAP and MELD datasets. During testing, Gaussian noise of varying intensities was injected into the textual, visual, and audio features, with the amplitude normalized by the overall feature standard deviation to simulate perception errors, environmental disturbances, and modality quality fluctuations. The experiments covered multiple levels, from a noise-free baseline to high-intensity noise (0.1–0.7), and used weighted accuracy and weighted average F1-score as evaluation metrics.

As shown in Fig. \ref{fig7}, model performance declined smoothly rather than fluctuating sharply with increasing noise intensity, indicating strong robustness. On IEMOCAP, when full-modality noise reached 0.3, weighted accuracy and average F1-score only slightly decreased compared to the baseline, and even at 0.7 noise, the performance degradation remained controlled. On MELD, even under noise levels of 0.5–0.7, weighted accuracy and average F1-score remained around 65\%, slightly lower than the baseline, demonstrating good noise resistance.

Further analysis indicates that DiffGCN consistently captures dynamic emotional dependencies both across modalities and within/between speakers. Its differential attention mechanism effectively reduces redundant inter-modality interference, preserving discriminative capability under noisy conditions. Meanwhile, the adaptive modality balancing mechanism dynamically adjusts modality weights under high-noise conditions, preventing dominant modalities from overwhelming the fusion process while retaining useful information from minor modalities, thus maintaining consistency and reliability of the model outputs under full-modality noise.

\subsection{Complexity Analysis} Table \ref{Table9}  summarizes the experimental results of the compared models in terms of parameter count and runtime. From the experimental data, although our model has a parameter scale similar to some graph convolutional models, its inference time is slightly higher, which is directly related to the introduction of the differential relational graph convolutional network and the adaptive modality balancing module. Experimental analysis shows that the differential relational graph convolutional network performs multiple attention computations and integrates relational embeddings for each node, enhancing the ability to capture dependencies both within and across speakers, but it also brings additional computational overhead. The adaptive modality balancing module dynamically adjusts the contribution of each modality in each batch, ensuring that weak modality information is effectively utilized; experiments show that its additional impact on overall inference time is relatively limited. Although there is some increase in computational cost, compared to the performance improvement, this overhead remains within an acceptable range. The experimental results indicate that the model achieves a good balance between high accuracy, computational complexity, and multimodal information fusion.

\begin{table}[htbp]
\centering
\caption{Model Parameters and Runtime Comparison}
\resizebox{\linewidth}{!}{
\begin{tabular}{l c c c c}
\toprule
\label{Table9}
\textbf{Method} & \textbf{Params (M)} & \textbf{IEMOCAP Runtime (s)} & \textbf{MELD Runtime (s)} \\
\midrule
DialogueGCN & 12.92 & 58.1 & 127.5 \\
RGAT & 15.28 & 68.5 & 146.3 \\
LR-GCN & 15.77 & 87.7 & 142.3 \\
MMGCN & 0.46 & 93.7 & 75.3 \\
DER-GCN & 78.59 & 125.5 & 189.7 \\
\hline
AMB-DSGDN (ours)& 13.13 & 179.95 & 146.4 \\
\bottomrule
\end{tabular}
}
\end{table}

\subsection{Adaptive Modality Balancing under Extreme Imbalance}

\begin{figure}[t]
	\includegraphics[width=0.99\linewidth]{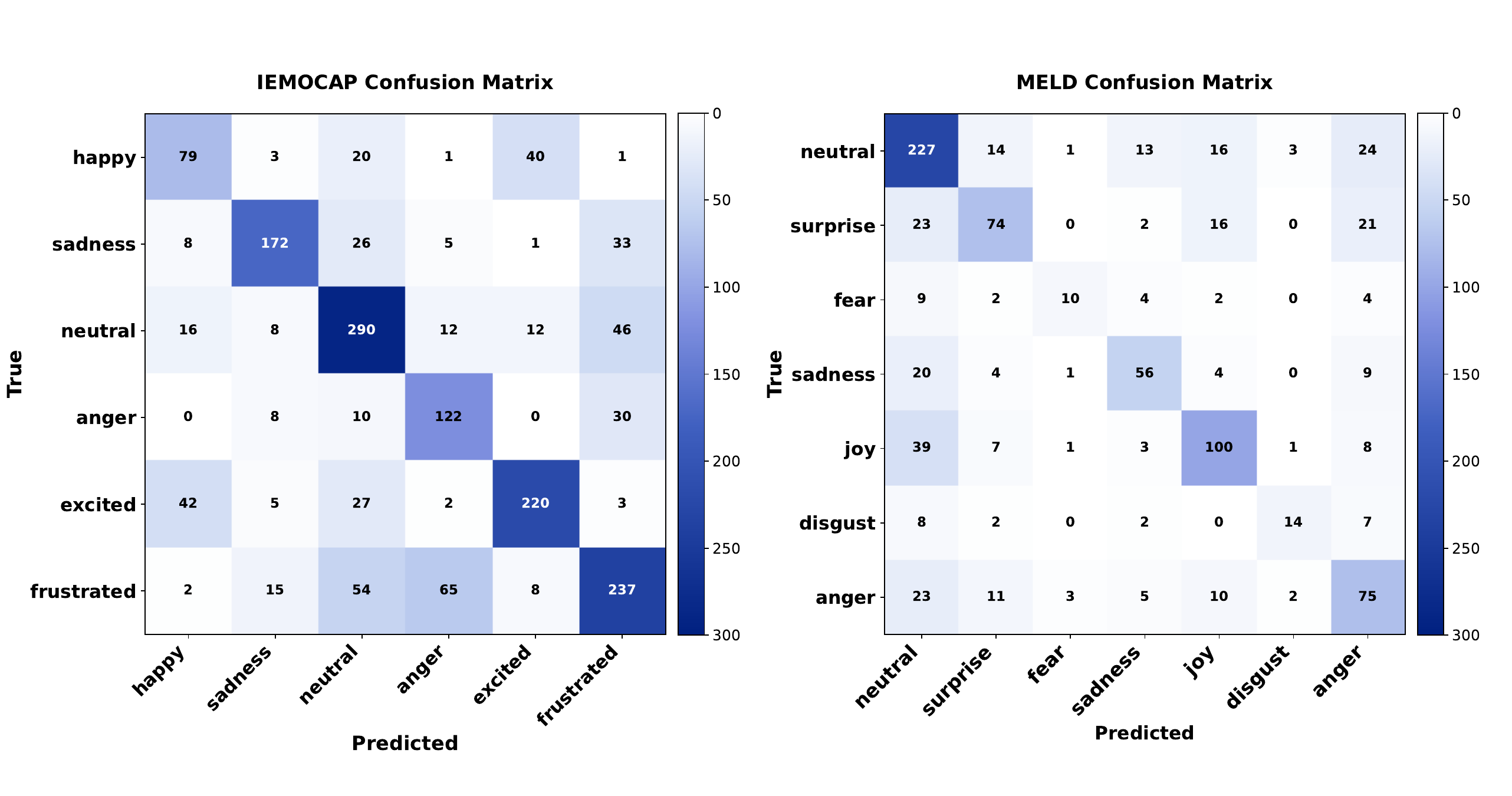}
	\caption{This figure presents the emotion recognition confusion matrices for the IEMOCAP (left) and MELD (right) datasets under the extreme modality imbalance experiment: the left matrix corresponds to the 6 emotion categories of IEMOCAP, while the right one corresponds to the 7 emotion categories of MELD. The "True" axis represents the ground-truth emotion labels, and the "Predicted" axis represents the model’s predicted labels; the values inside the cells indicate the number of samples for the corresponding category. The dark-colored blocks along the matrix diagonal reflect the high recognition accuracy for most emotions, while the light-colored blocks off the diagonal reflect the misclassification between different emotions.}
	\label{fig8}
\end{figure}

Based on the single-modality performance differences in Table \ref{Table4}, we conducted extreme modal imbalance experiments on the IEMOCAP and MELD datasets, setting the fusion weights for text, visual, and audio modalities to 0.8, 0.1, and 0.1, respectively. Results in Fig. 
\ref{fig8} show that, despite the dominance of the text modality, the model maintains high overall accuracy and weighted F1 scores. Further analysis of the confusion matrices indicates that the visual and audio modalities still make significant contributions in recognizing key emotions such as “anger” and “excited,” effectively mitigating misclassifications caused by text modality bias. This demonstrates that the model can adaptively regulate the information flow based on the relative effectiveness of each modality, preventing the dominant modality from overwhelming the fusion result. Combined with the differential graph attention mechanism, this regulation also suppresses shared noise, ensuring a more balanced contribution from each modality at different stages. Overall, the experiments validate the effectiveness of the adaptive modality balancing mechanism in enhancing model robustness and sensitivity to dynamic multimodal emotional variations under extreme modal imbalance.

\section{Conclusion}
In this study, we propose AMB-DSGDN for multimodal conversation emotion recognition. The model effectively mitigates modality contribution imbalances through an adaptive modality dropout mechanism based on real-time performance evaluation. Meanwhile, it independently constructs inter-speaker and intra-speaker relational subgraphs for each modality and incorporates a DiffRGCN, which captures the dynamic evolution of emotional dependencies via positive-negative subspace projections and differential attention operations. Experimental results demonstrate that AMB-DSGDN significantly outperforms existing baseline methods on the IEMOCAP and MELD datasets, achieving stable improvements in overall performance and specific emotion categories. 

However, due to the computational complexity and overhead introduced by graph-structured modeling, the inference efficiency of the proposed model remains limited when processing extremely long dialogue sequences and operating in resource-constrained environments. In particular, for deployment on edge devices, challenges such as model size, the computational cost of graph attention mechanisms, and the real-time processing of multimodal features need to be carefully addressed. Future work will focus on improving computational efficiency by exploring lightweight graph attention designs, subgraph pruning, and parameter sharing strategies, as well as investigating model compression, knowledge distillation, and hardware-aware acceleration techniques to enhance real-time inference performance on edge devices. These efforts are expected to further improve the applicability of the proposed framework in real-time human–machine interaction and emotion analysis scenarios.

\section*{CRediT authorship contribution statement}

\textbf{Yunsheng Wang}: Conceptualization, Methodology, Investigation, Data curation, Writing - Original Draft. \textbf{Yuntao Shou}: Supervision, Investigation \& Review. \textbf{Yilong Tan}: Supervision, Investigation, Writing - Review \& Editing. \textbf{Tao Meng}: Supervision, Investigation, Writing - Review \& Editing. \textbf{Wei Ai}: Supervision, Investigation, Writing- Review \& Editing. \textbf{Keqin Li}: Supervision, Investigation, Writing - Review \& Editing.

\section*{Declaration of Competing Interest}

The authors declare that they have no known competing financial interests or personal relationships that could have appeared to influence the work reported in this paper.

\section*{Data availability}

Data will be made available on request.

\section*{Acknowledgements}

The authors deepest gratitude goes to the anonymous reviewers and AE for their careful work and thoughtful suggestions that have helped improve this paper substantially. This
work is supported by National Natural Science Foundation
of China (Grant No. 69189338), Excellent Young Scholars
of Hunan Province of China (Grant No. 22B0275), and program of Research on Local Community Structure Detection Algorithms in Complex Networks (Grant No. 2020YJ009).

\bibliographystyle{ieeetr}
\bibliography{reference}
\end{document}